\documentstyle[preprint,aps,eqsecnum,epsfig]{revtex}
\tightenlines
\begin{document}
\draft 
\title{\bf DYNAMICS OF SYMMETRY BREAKING OUT OF EQUILIBRIUM: FROM CONDENSED MATTER TO QCD 
AND THE EARLY UNIVERSE\footnote{Invited by the National Academy of Sciences of India, to appear 
in ANSI-2000, and in the Proceedings of the VI\`eme Colloque Cosmologie,
 Paris, 16-18 JUNE 1999 } } 
\author{{\bf D. Boyanovsky$^{(a,b)}$ and  H.J. de Vega$^{(b,a)}$}}  
\address { (a)Department of Physics and Astronomy,
University of Pittsburgh, Pittsburgh, PA 15260 USA\\ (b)
LPTHE\footnote{Laboratoire Associ\'{e} au CNRS UMR 7589.}  Universit\'e Pierre
et Marie Curie (Paris VI) et Denis Diderot (Paris VII), Tour 16, 1er. \'etage,
4, Place Jussieu 75252 Paris, Cedex 05, France} 
\date{\today} 
\maketitle
\begin{abstract}
The dynamics of symmetry breaking during out of equilibrium phase transitions is a topic of great importance 
in many disciplines, from condensed matter to particle physics and early Universe cosmology with definite 
experimental impact.  In these notes we provide a summary of the relevant aspects of the dynamics of symmetry
 breaking in many different fields with emphasis on the experimental realizations. In condensed matter we address
 the dynamics of phase ordering, the emergence of condensates, coarsening and dynamical scaling. In QCD the possibility
 of disoriented chiral condensates of pions emerging during a strongly out of equilibrium phase transition is discussed.
We elaborate on the dynamics of phase ordering in phase transitions in the Early Universe, in particular the 
emergence of condensates and scaling in FRW cosmologies. We mention some experimental efforts in different fields that 
study this wide ranging phenomena and offer a quantitative theoretical
description both at the phenomenological level in condensed matter, introducing the scaling hypothesis as well as at
 a microscopic level in quantum field theories. The emergence of semiclassical condensates and a dynamical length scale 
is shown in detail, in quantum field theory this length scale is constrained by causality.

The large N limit provides a natural bridge to compare the solutions in  different settings and to establish 
similarities and differences. 

\end{abstract}

\section{Phase Ordering Dynamics: an interdisciplinary fascinating problem}
The dynamics of non-equilibrium phase transitions and the ordering
process that occurs until the system reaches a 
broken symmetry equilibrium state play an important role in many
different areas. In condensed matter physics  binary 
fluids, ferromagnets, superfluids, and liquid crystals, to name a few, are examples of systems in which the 
dynamics of phase transitions out of equilibrium are studied experimentally. 

Experiments in these systems have provided a solid basis for the description of the dynamics of phase ordering: 
in binary fluids or alloys upon a sudden temperature drop below the critical temperature, the two fluids begin
to separate, regions of different fluid concentrations are separated by
{\em domain walls}. In superfluids, rapid cooling leads to a network of
vortices and in liquid crystals to many different topological defects.  

 In cosmology defects are conjectured to be produced 
during Grand Unified Theory (GUT) or the Electro-weak 
(EW) phase transition
can act as seeds for the formation  of large scale structure 
and the dynamics of phase ordering and formation of
ordered regions is at the heart of Kibble's mechanism of 
defect formation\cite{kibble,kibble2,vilen}. Current and
future measurements of Cosmic Microwave Background 
anisotropies  could
provide distinct evidence for these phase transitions whose dynamics could have influenced structure 
formation\cite{durrer}. 
At even lower energies, available with current
and forthcoming accelerators, the Relativistic Heavy Ion Collider (RHIC) at Brookhaven  and the Large Hadron 
Collider (LHC) at Cern the phase transitions
predicted by the theory of strong interactions, Quantum Chromodynamics
(QCD) could occur out of equilibrium resulting in the formation of coherent
condensates of low energy Pions. These conjectured configurations
known as `Disoriented Chiral Condensates'
are similar to the defects expected in liquid crystals or ferromagnets in condensed matter systems 
and their charge distribution could be an  
experimental telltale of the chiral phase transition of 
QCD\cite{rajagopal}. Whereas the GUT phase transition took
place when the Universe was 
about $10^{-35}$ seconds old and the temperature about $10^{23}K$, and
the EW phase transition occured when the 
Universe was $10^{-12}$ seconds old and with a temperature $10^{15}K$,
the QCD phase transition took place at about 
$10^{-5}$ seconds after the Big Bang, when the temperature was a mere
$10^{12}K $. This temperature range will be probed 
at RHIC and LHC within the next very few years. The basic problem of
describing the process of phase ordering, the 
competition between different broken symmetry states  and the formation and evolution of condensates on the way towards  
reaching equilibrium is common to all of these situations and fields. The tools,
however, are necessarily very different: whereas 
ferromagnets, binary fluids or alloys etc, can be described via a
phenomenological (stochastic) classical description, certainly in quantum 
field theory a microscopic formulation must be provided. In these lectures
we describe  a program to include 
 ideas from condensed matter to the realm of quantum field theory, to
describe the non-equilibrium dynamics of symmetry breaking and the process of phase separation and phase ordering 
on a range of time and spatial 
scales of unprecedented resolution (for example in QCD the time scales $\leq  10^{-23}$
seconds, spatial scales $\leq 10^{-15}$ meters, in cosmology the time scales
are of order $10^{-32}$ seconds and spatial scales smaller than $10^{-44}$ meters) that 
require a full quantum field theoretical  description. 

We begin the excursion into these timely fields by first providing a brief
quantitative description of the relevant setting and whenever possible 
the experimental situation associated with them in three main areas:
Condensed Matter, Ultrarelativistic Heavy Ion Collisions and Early Universe Cosmology. This quantitative 
discussion will be followed by a more qualitative description of some of the main theoretical ideas, techniques 
and tools. 

\subsection{Condensed Matter:}
A description of phase transitions and critical phenomena in equilibrium
begins by recognizing an {\em order parameter} which is a thermodynamic
ensemble average of a macroscopic variable that determines the different
macroscopic states of the system. For example in ferromagnets the order
parameter is the average magnetization, above a critical temperature it
vanishes and it is non-zero below the critical temperature, in superfluids
is the condensate density, in superconductors the density of Cooper pairs,
etc.\cite{goldenfeld}. Phase transitions {\em in equilibrium} are fairly
well understood and described by the theory of critical phenomena\cite{goldenfeld} which combined 
with the renormalization group provides a very succesful description of phase transitions. The theory
of critical phenomena and the renormalization group provide a very robust 
description of {\em universality classes}: many systems that are very 
different behave similarly near critical points, these universality
classes are divided by for example the dimensionality of the order parameter, the dimensionality of space, 
and the symmetries of the underlying microscopic Hamiltonian. An important concept in critical phenomena is
 the
correlation length, take for example a spin one-half ferromagnet, the
microscopic Hamiltonian has an up-down symmetry, the energy remains the same if all spins are flipped. Focus 
at a particular point of the sample where the spin is  up. The correlation length is the distance over which 
the spins are correlated, i.e. the distance from this up-spin over which the neighboring spins are also up. As 
the critical
temperature is reached from above this correlation length grows reaching
a macroscopic size (diverging) at the critical temperature. As the system
is cooled below the critical temperature a phase transition occurs: there
appears a net overall magnetization and at low temperatures all spins are
either up or down, the up-down symmetry is spontaneously broken\cite{goldenfeld}. This phase transition
 occurs in equilibrium when
the microscopic relaxation time scales are shorter than the time scale of
cooling the system, thus at all times the system is in {\em local thermodynamic equilibrium}. At very high 
temperatures typically the 
disordered phase prevails, all spins are oriented at random and the average magnetization vanishes. As the 
critical temperature is reached regions of correlated spins appear and become of macroscopic size as the 
correlation
length diverges and the spin system begins to order. In this region the thermodynamic quantities become 
insensitive to the short distance details such as crystalline lattices, lattice spacing and the nature and strength 
of the interaction between
spins as the physics is determined by the correlation of spins over large
distances. Near the critical point, the short distance length scales are
irrelevant for macroscopic phenomena and long-wavelength physics is completely determined by the correlation 
length $\xi(T)$. Macroscopic thermodynamic quantities and susceptibilities near the critical temperature
{\em only} depend on the length scale $\xi(T)$.

 This is the basis of the {\em static} scaling hypothesis which is confirmed experimentally in a wide variety
 of systems and is theoretically  supported by the renormalization group approach to critical phenomena\cite{goldenfeld}. 
The {\em static} critical phenomena associated with 
second order phase transitions that occur in local thermodynamic equilibrium is fairly well understood via the
 renormalization group (and other alternative approaches)\cite{goldenfeld}. 

Consider the alternative scenario in which a ferromagnet is held at very
high temperature in the disordered phase and suddenly it is cooled below
the critical temperature on time scales shorter than those associated with
relaxational phenomena. Now the spin system must evolve towards the ordered
phase far away from equilibrium. Unlike the case of static (local thermodynamic equilibrium) 
critical phenomena, the case of out of equilibrium phase transitions  require a novel set of ideas and tools to
describe the {\em dynamics} of the process of phase ordering.  

There is now  a large
body of theoretical and experimental work in phase ordering dynamics in
condensed matter systems\cite{bray}-\cite{marco}. Although ultimately
the tools to study similar questions in  quantum mechanical many body systems will be  different, the 
main physical features to describe are basically the same:
as the system cools down suddenly below the critical temperature correlated
regions (of spins in a ferromagnet or of condensate in a Bose superfluid)
begin to form. These  correlated regions are  separated 
by `walls' or other structures. Inside these regions an ordered phase
exists which eventually grows in time to 
become macroscopic in size. 

Before attempting to describe the manner
in which a given system orders after being cooled 
through a phase transition an understanding of the relevant time
scales is required. Two important time scales determine 
if the transition occurs in or out of equilibrium: the relaxation time
of long wavelength fluctuations (since these 
are the ones that order) $\tau_{rel}(k)$ and the inverse of the
cooling rate $t_{cool}= T(t)/\dot{T}(t)$. 
If $\tau_{rel}(k)<<t_{cool}$ then these wavelengths are in local
thermodynamical equilibrium (LTE), but if  
$ \tau_{rel}(k)>>t_{cool} $ these wavelengths fall out of LTE and freeze out, for
these the phase transition occurs 
in a quenched manner. These modes do not have time to adjust
locally to the temperature change and for them the 
transition from a high temperature phase to a low temperature one
occur instantaneously. This description was 
presented by Zurek\cite{zurek}  analysing   the emergence of defect
networks after a quenched phase transition.   
Whereas the
short wavelength modes are rapidly thermalized (typically by
collisions) the long-wavelength modes with $ k << 1/\xi(T) $ with
$ \xi(T) $ the correlation length (in the disordered phase) become {\em
critically slowed down} i.e. their relaxation time becomes extremely long near the critical point. 
As $ T\rightarrow T_c^+ $ the long 
wavelength modes relax very slowly, they fall out of LTE and any
finite cooling rate causes them to undergo a ``quenched'' non-equilibrium
phase transition. As the system is quenched from $ T>T_c $ (disordered 
phase) to $ T<<T_c $ (ordered phase) ordering {\em does not} occur
instantaneously. The length scale of the ordered 
regions grows in time (after some initial transients) as the different
broken symmetry phases compete to select the 
final equilibrium state. A {\em dynamical} length scale $ \xi(t) $ typically
emerges  which is interpreted as the size of 
the correlated regions, this dynamical correlation length grows in
time to become macroscopically large\cite{bray,langer,mazenko,marco}. 
Just as in {\em static} critical phenomena, the emergence of this dynamical correlation length leads to the 
{\em dynamical scaling hypothesis}, that the
approach to equilibrium and the kinetics of phase ordering is solely determined by this length scale.

Experiments in binary fluids for example, study the growth of these
correlated regions by light scattering\cite{goldburg} much in the same
manner as the onset of ferromagnetism is studied via neutron scattering. The
growth of the domains is characterized by the dynamical length scale. 
As a function of time the scattering of light becomes stronger
for longer wavelengths i.e. smaller wave-vectors, until eventually at
very long times a Bragg peak at zero momentum emerges signaling the macroscopic ordering of the system. 
This growth of domain structures during the dynamical process of phase ordering is referred to as 
``coarsening''\cite{bray,langer,mazenko,marco}. This mechanism with a
clear experimental realization in condensed matter is at the heart of
the Kibble-Zurek\cite{kibble,kibble2,zurek,vilen,gill} scenario of the dynamics of symmetry breaking 
in cosmological phase transitions\cite{kibble,kibble2,zurek,vilen,gill}. In this scenario a ``network''
 of defects emerges  after a phase transition that occurred
strongly out of equilibrium with a density of about one defect per initial
correlation length. Computer simulations reveal that this network  evolves and  a scaling regime  emerges\cite{vilen}.

This idea
had sparked an intense effort to reproduce ``cosmological phase transitions in the laboratory''.
The dynamics of phase ordering had been studied
in liquid crystals whose symmetry group is rather similar to that of particle physics models. The 
experiments produced non-equilibrium phase transitions both by suddenly varying the pressure and the 
temperature (pressure and temperature quenches)\cite{turokexp,bowickexp}  and confirmed at least in a 
qualitative
manner the main features described by this scenario of dynamics of symmetry breaking. More recently a new set of
 experiments had sought to provide
a more detailed picture of the dynamics of symmetry breaking phase transitions out of equilibrium and to simulate
 in the laboratory what is thought to be the situation in cosmological phase transitions. 
Original experiments focused on studying the dynamics of phase ordering
after a pressure quench in superfluid $^4\mbox{He}$\cite{lancaster} by
measuring second sound (i.e. entropy disturbances) which only propagate in
the superfluid component (the broken symmetry phase). The interpretation of results in these experiments
 were overshadowed by induced turbulence during the quench and spurious phase separation due to imperfections
 of the walls.
More recently a new set of experiments were carried out that seem to lead to
cleaner interpretations. 

In these
ingenious experiments\cite{ruutu,bunkov} a small sample of superfluid
$^3\mbox{He}$, whose order parameter has a group structure very similar to
some particle physics models, 
 was heated locally by neutron irradiation via the
nuclear reaction
$$
^3\mbox{He}+ n = ^3\mbox{He}^- + p + \mbox{764 keV}
$$
the energy released heats a small portion of the liquid Helium into the
normal state and rapid diffusion of the quasiparticles cools this region
back into the superfluid phase very rapidly, thus providing a quench from
a normal (disordered) phase into the superfluid (ordered) phase. The
resulting domain structure is then studied via NMR and a qualitative
agreement with the picture of the symmetry breaking dynamics seem to emerge from these experiments. 
 Thus these beautiful experiments in condensed matter reproduce in a qualitative manner the physics of
 a ``little Big Bang'' and provide controlled experimental framework to test the concepts associated
with the dynamics of symmetry breaking.

These ideas of the emergence of correlated regions that grow in time and
become macroscopic during non-equilibrium phase transitions has been recently invoked as a potential
 signature of the chiral phase transition
in QCD, the theory of the strong interactions.

\subsection{Chiral symmetry breaking in QCD and
 ``disoriented chiral condensates'' }

Quantum Chromodynamics (QCD) is the theory of strong interactions, with the
fundamental degrees of freedom being the quarks and gluons. Quarks, however
are confined inside hadrons and do not exist as individual, isolated particles in vacuum. However
 there is now a wealth of theoretical evidence
including very convincing lattice results that indicate that at temperatures
above $T \approx 150 \mbox{Mev}$ quarks and gluons become free and form
a quark-gluon plasma. The lattice results are supported qualitatively and quantitatively by phenomenological
models\cite{QCD}. In fact the numerical evidence supports the picture of {\em two}
phase transitions: the quark-gluon plasma or confining-deconfining phase
transition in which the quarks and gluons become confined into hadrons and the chiral phase transition that leads 
to the low energy description in terms of pions. The low energy limit of QCD
is dominated by the lightest up and down  quarks $u,d$ with masses $m_u \approx 5 \mbox{Mev}; m_d \approx 7-10 \mbox{Mev}$. 
These mass scales are much smaller
than the natural scale of QCD, $\Lambda_{QCD} \approx 100 \mbox{Mev}$ at which QCD becomes strongly coupled. In
 the limit of vanishing up and down
quark masses, the QCD Hamiltonian possesses a global {\em chiral symmetry} corresponding to rotating independently
 the right and left handed components of the spinors that describe the quark fields. This symmetry is $SU(2)_L \otimes SU(2)_R$ and in the low energy world is spontaneously broken
down to $SU(2)_{L+R}$ with the charged and neutral pion isospin triplet being the (quasi) Goldstone bosons associated
 with the breakdown of this symmetry. The small mass of the pions
 ($\approx 135 \mbox{Mev}$), on the hadronic scale is a result of the small mass of the up and down quarks on the 
QCD scale, which breaks explicitly chiral symmetry.
This is the chiral phase transition. The lattice results seem to indicate that the two transitions, deconfining and
 chiral symmetry breaking are very
close in temperature and may actually happen at the same temperature\cite{QCD}.

Whereas the deconfining phase transition does not seem to be characterized
by a natural order parameter, the chiral transition is described by the
non-vanishing of the chiral condensate $<\bar{q}q>$ with $\bar q = (\bar u, \bar d)$ with $<\cdots >$
 refering to the vacuum expectation value or the thermodynamic ensemble average 
at finite temperature. Although this transition(s) have taken place when the Universe was at a temperature
of $150 \mbox{Mev}$ about $10^{-5}$ seconds after the Big Bang, the Relativistic Heavy Ion Collider (RHIC) at
 Brookhaven to start operation
at the end of 1999 and the Large Hadron Collider at Cern (around 2004) will
probe this transitions by colliding heavy ions. 

RHIC will accelerate and collide from protons up to 250 Gev and ions of up to the heaviest nuclei with 
collision energies of about 100 Gev per nucleon
for Au nuclei. The phenomenon of nuclear transparency observed in nucleon-nucleon collisions leads to 
the conclusion that about half the energy of
the collision is carried away by the nuclei and about half the energy is
deposited in the ``central region'' of the collision. Most of the baryons
are carried by the receding nuclei (the fragmentation region)
 leaving this central region almost baryon free. Estimates of the energy density 
in this region give $\epsilon \approx 
3-5 \mbox{Gev}/\mbox{fm}^3$ corresponding to temperatures $T\approx 200 \mbox{Mev}$. Immediately after the collision, 
hard scattering of quarks and gluons dominate the dynamics the gluons have mean-free paths estimated
to be of order $0.5 \mbox{fm}$ and the quarks $1-2 \mbox{fm}$ (the difference is mainly due to color factors) hence after
  a time of the order of
about $1\mbox{fm}/c$ the plasma is thermalized. 

 The next stage of the
dynamics is described by Bjorken's hydrodynamic picture\cite{bjorkenhydro}. When the plasma has
achieved local thermodynamic equilibrium and for wavelengths longer than the
mean free paths, the plasma can be described as a strongly coupled fluid and a hydrodynamic description is 
suitable. The essential ingredients in a hydrodynamics description are i) the fluid is described by a {\em local}
 four velocity vector $u^{\mu}= \gamma (1,\vec v); u^{\mu}u_{\mu}=1$, the
energy momentum tensor is that of an homogeneous and isotropic fluid

$$
T^{\mu \nu} = (\epsilon + P)u^{\mu} u^{\nu} - P g^{\mu \nu}
$$ 
with $\epsilon ~, ~ P$ the energy density and pressure respectively. The dynamics is then completely determined by 
conservation laws: i)  baryon
number, ii) energy momentum and by local thermodyamic equilibrium relations. The
resulting picture of this hydrodynamic evolution is that the plasma expands and cools adiabatically and the 
temperature drops in time with the following
law
$$
T(\tau) = T_0 \left(\frac{\tau_0}{\tau} \right)^{c^2_s}
$$ 
with $c_s$ the adiabatic sound speed, $T_0 \geq 200 \mbox{Mev}$ and $\tau_0 \approx 1\mbox{fm}/\mbox{c}$. For
 a radiation dominated fluid $c^2_s=1/3$.  

As the critical temperature for the chiral phase transition is reached
from above the long-wavelength fluctuations of the chiral order parameter
are expected to become critically slowed down. If this is the case the
chiral phase transition can occur in a ``quenched manner'' and strongly
out of equilibrium. Under these circumstances, Wilczek and Rajagopal
argued that large domains in which the chiral order parameter could be
``disoriented'' with respect to the vacuum could appear\cite{raja}. These
domains are coherent pion condensates that form after the non-equilibrium phase transition much in the
 same manner as the correlated domains in condensed matter systems. These pion condensates  decay, the neutral pion
decays into two photons and the charged pions decay into muons. The pions
can then be reconstructed and therefore these disoriented chiral condensates
could lead to experimentally observable anomalies in the ratio of the
number of neutral to charged pions. In isospin symmetric condensates
the probability for finding a ratio R of neutral to total (neutral plus charged) is $P(R) \propto \delta(R-1/3)$ 
(for large number of pions) whereas
a disoriented chiral condensate leads to $P(R)\propto 1/\sqrt{R}$\cite{dcc}.

 The possibility of formation of disoriented chiral condensates had been previously conjectured by 
Bjorken\cite{dcc} 
as a potential explanation of CENTAURO events\cite{centauro}, these are cosmic rays events
with anomalous neutral to charged pion ratios. This possibility of a distinct signature associated with the chiral phase transition sparked an intense
 theoretical effort\cite{moredcc,boydcc}. 
Several experimental searches
are trying to find evidence for this pion condensates or disoriented
chiral condensates at CERN-SPS (experiment WA98\cite{wa98}) at the Tevatron at Fermilab (Minimax experiment\cite{fermilab}), the PHENIX and STAR detectors
at RHIC\cite{rhic} at BNL can provide an event-by-event analysis of this potential candidates
and the ALICE experiment scheduled at CERN-LHC includes the detector
 CASTOR\cite{castor} 
that will be studying CENTAURO type events.

These disoriented chiral condensates are coherent pion domains and describe the same type of phenomenon as 
 domains in quenched ferromagnets or 
 in He superfluids as described previously. If these condensates are realized during a
 non-equilibrium stage of the chiral phase transition, they  could lead to important experimental probes of this transition and 
hopefully will be amenable of detection at the RHIC and LHC on an event by event basis.

There is an important difference in the dynamics of the chiral phase transition in the Early Universe and at Ultrarelativistic Heavy Ion Colliders. In the Early Universe, the chiral phase transition occured at
a temperature of $150 \mbox{Mev}$ when the Universe was about $10^{-5}$ seconds old in the radiation dominated era. The size of the Universe
at that time was about 10 Km which is much larger  than the mean-free path of quarks and gluons $\approx 10^{-15}\mbox{m}$ and the time scale for cooling $T/\dot{T}\approx 10^{-5}\mbox{secs}$ is {\em} much longer 
than the relaxation time scale of partons $\tau_{rel} \approx 10^{-23}\mbox{secs}$. Therefore
in the Early Universe the confining and chiral phase transition most likely occured in
{\em equilibrium}. These time and length scales must be compared to those
in heavy ion collisions: the time scale for cooling from hydrodynamic expansion is few fm/c and the relaxation time scale near phase transitions
could be longer and comparable to the lifetime of the quark-gluon plasma, furthermore current numerical estimates determine that the 
region in which the QGP is formed is about 20 fm. Hence there is a possibility that these phase transitions could be out of equilibrium in heavy ion collisions and that novel phenomena associated with the process of phase ordering and the emergence of pion 
condensates could be important experimental signatures of the chiral transition.

\subsection{Early Universe Cosmology:} The COBE satellite mission revolutionized
the field of Cosmology. The discovery of temperature fluctuations in
the Cosmic Microwave Background (CMB) of $30 \mu K$ imprinted on a
blackbody spectrum at $2.73 K$ provides supporting evidence for the main
ideas that seek to explain  the small inhomogeneities that gave
rise to large scale structure formation\cite{kolb,turner}.  Two leading
contenders with a solid base on particle physics have emerged that provide
different explanations for the origin of the primordial inhomogeneities that grew via gravitational instability to form large scale structures: inflation and defects\cite{kolb,turner,durrer}. Inflation postulates that at an energy
scale determined by Grand Unified Theories $\approx 10^{16}\mbox{Gev}$
the Universe underwent a period of exponentially accelerated expansion during which its size  grew by a factor $e^{60}$ necessary
to solve several problems with the standard Big Bang Cosmology\cite{kolb,turner}. Small quantum fluctuations that were present
during this epoch of inflation soon became causally disconnected and therefore unaffected by microphysical processes. These fluctuations became
in causal contact again at a much later stage of the cosmological evolution,
when the Universe was basically dominated by matter. Small fluctuations begin to grow under gravitational instability when they become causally connected again
but after the epoch of radiation-matter equality at a temperature of about
$10 \mbox{eV}$ and redshift $z \approx 10^4$\cite{kolb}.
The COBE experiments
are sensitive to those fluctuations that have established causal contact
again after the epoch of recombination about 300000 years after the Big Bang at a redshift of about $z \approx 1100$. Therefore observations of the CMB allow to obtain information on the spectrum of primordial {\em quantum} fluctuations that were present shortly after the Big Bang.  

Original scenarios of inflation
relied on a supercooled phase transition\cite{kolb}. Recent detailed studies of the {\em dynamics} of phase transitions in early universe cosmology\cite{inflation} allow
a reliable calculation of the dynamics including backreaction effects on the metric and a self-consistent evolution of classical gravity and quantum fields. This approach allows to extract  the power spectrum of the primordial perturbations of the
 metric arising from the quantum fluctuations in the matter fields. These fluctuations  are directly related to those of the temperature of the CMB at the scale of recombination and correspond to the Sachs-Wolff plateau in the power spectrum measured by COBE\cite{kolb,turner}. It 
is found\cite{inflation} that the growth of correlated domains after a supercooled
phase transition of second order (no metastability) favors  a power spectrum with more power on long wavelengths\cite{inflation} as a consequence of the
process of phase ordering. This is a consequence of the unstabilities associated with the early stages of the phase ordering dynamics.  This ``red'' power spectrum is consistent with
the results of COBE for the temperature anisotropies, provided that the couplings of the matter field are fine tuned\cite{inflation}. 

The alternative proposal for explaining the small primordial 
fluctuations that result in the small temperature inhomogeneities observed
on the CMB invoke the formation and subsequent evolution of a network of
defects after a cosmological phase transition out of equilibrium\cite{kolb,turner}. In these scenarios, the dynamics of phase ordering after a phase transition results in a network of ``cosmic strings''\cite{kibble,kibble2,vilen,durrer}. The phase transitions required in these scenarios occur at a GUT scale $\approx 10^{16}\mbox{Gev}$ and the
important quantity is the string tension\cite{vilen} which is the energy per unit length of these topological defects and which determines the
amplitude of the fluctuations on the space-time metric. There is a very
important difference between the fluctuations in the inflationary and the
topological defects scenarios. In inflation, the quantum fluctuations become causally disconnected\cite{kolb,turner} and therefore their evolution is
very simple until they become causally connected again because these
fluctuations are not influenced by microphysical processes during the period of acausal evolution. Contrary to this
``acausal'' dynamics, topological defects are always causal and are constantly influenced by microphysical processes. Their evolution must
be followed dynamically from the time at which the network of defects is formed, at
a GUT scale, all the way up to the time scale at which they result in the
formation of large scale structure--several billion years later!. Obviously this is an enormous dynamical range, however, detailed computer simulations reveal that a {\em scaling}
solution emerges (for details see\cite{vilen}) determined by a dynamical
length scale. The results of numerical studies suggest that this dynamical
length scale is completely determined by the size of the causal horizon
at a given time (see the later section on Cosmology for details on causal horizons).  

The emergence of this length scale through the dynamical process of phase ordering is exactly the same that has been 
 previously discussed within the context of condensed matter systems. Current ground based and balloon borne 
experiments along with large scale surveys and future satellite missions will provide a flood of data 
that will support or falsify current theoretical ideas on large scale structure formation and temperature anisotropies. 
Thus an important theoretical effort goes in providing reliable predictions on the power spectrum of primordial quantum 
fluctuations. It is a tantalizing possibility that these cosmological observations could provide a definite evidence for 
cosmological phase transitions. 

\subsection{...Therefore} We have seen in detail that the dynamics of phase ordering and evolution after
 non-equilibrium phase transitions are of fundamental importance in a wide range of energies from meV, in Condensed Matter, through Gev in the physics of the Quark Gluon Plasma and the Chiral Phase Transition all the way to GUT's ($10^{16}\mbox{Gev}$) in Early Universe Cosmology. 
An important technical aspect in the study of these
phenomena is their {\em non-perturbative} nature: in a rapid phase transition (of typical second order without metastable states) small amplitude long-wavelength fluctuations become unstable (this will be understood in detail below) and grow in time. The amplitude of these fluctuations must grow until they begin to sample the broken symmetry states of thermodynamic equilibrium.

Furthermore, in quantum field theory the notion of a ``defect'' requires a careful quantitative understanding of the
process of {\em classicalization and decoherence} of quantum fluctuations. Defects are intrinsically classical objects and
are typically described as arising in a field theory as solutions of the classical equations of motion. Whether the
description of the dynamics of quantum phase transitions in terms of the emergence of defects or coherent structures is
a {\em correct or appropriate one} can only be determined by following the time evolution of the quantum density matrix.
The classicalization of fluctuations and the emergence of semiclassical coherent, large amplitude field configurations
should be a consequence of the time evolution and not an {\em a priori} description of the dynamics. Therefore one of our
main goals is to provide a consistent quantum field theory description of the dynamics that is capable of describing
this process of classicalization and decoherence of fluctuations.

Having discussed in some detail the importance of the dynamics of symmetry breaking phase transitions out of equilibrium within important settings and their experimental study, we now provide some of the technical aspects that help clarify the phenomena and their quantitative study. We begin by describing a phenomenological approach to phase ordering kinetics in condensed matter systems, highlighting the important ingredients and concepts. We then move on to furnish a quantitative approach to the study of the non-equilibrium dynamics in quantum field theory to compare some striking similarities to condensed matter and also to contrast some important and relevant differences. The main point for delving into some technical details is to emphasize many {\em robust} features of the dynamics of symmetry breaking and phase ordering, 
\begin{itemize}
\item {The early stages of phase ordering are determined by linear (spinodal) instabilities. Long-wavelength fluctuations become unstable and grow.}

\item {The emergence of a {\em dynamical} length scale. This scale represents the average size of the ordered domains and grows in time, eventually at asymptotically long times becoming macroscopically large.}

\item {Associated with this dynamical length scale there is {\em dynamical scaling}, asymptotically this length scale determines the behavior of correlation functions. }

\item {Coarsening: the growth in time of this correlation length translates in that the peak of the power spectrum moves towards longer wavelength, resulting in a sharp ``Bragg peak'' at asymptotically long times. This Bragg peak reflects the onset of  condensates corresponding to ordered regions of macroscopic size. }
\end{itemize}
 
As we will see in detail, the phenomenological description in condensed matter systems is {\em very different} from the microscopic description in quantum field theory. Nevertheless we find that despite these important differences the above features are fairly robust and common to all of the
situations studied. Only an excursion into the technical details can reveal in full force these very important and remarkable features. 

\section{Phenomenology of phase ordering dynamics in Condensed Matter:}
The phenomenological  description of phase ordering kinetics begins
with a coarse grained local free energy functional of 
a (coarse grained) local order parameter $M(\vec r)$\cite{bray,langer}
which determines the {\em equilibrium} states. In Ising-like systems
this $M(\vec r)$ is the local magnetization 
(averaged over many lattice sites), in binary fluids or alloys
it is the local concentration difference, in superconductors is the
local gap, in superfluids is the condensate fraction etc. The typical free energy is (phenomenologically) of the Landau-Ginzburg form:
\begin{eqnarray}
F[M] & = & \int d^d\vec x \left\{\frac{1}{2} [\nabla M(\vec x)]^2 +
V[M(\vec x)]\right\} \nonumber \\ 
V[M] & = & \frac{1}{2} \;  r(T)\,  M^2 + \frac{\lambda}{4}\,M^4  ~~; ~~ r(T) =
r_0(T-T_c) \label{freenergy} 
\end{eqnarray} 

Fig. 1 depicts $V[M]$ for $T>T_c$ and $T<T_c$. 
The equilibrium states for $T<T_c$ correspond to the broken symmetry states with $M= \pm M_0(T)$ with

\begin{equation}
M_0(T) = \left\{ \begin{array}{cc}
0 & ~\mbox{for} ~ T>T_c \\
\sqrt{\frac{r_0}{\lambda}}(T_c-T)^{\frac{1}{2}} &  ~\mbox{for} ~T<T_c
\end{array} 
\right. \label{MofT}
\end{equation}
 
Below the critical temperature the potential $V[M]$ features a non-convex
 region with $\partial^2 V[M] / \partial M^2 <0$  for
\begin{equation}
-M_s(T)<M<M_s(T) ~~; ~~ M_s(T) = 
\sqrt{\frac{r_0}{3\lambda}}(T-T_c)^{\frac{1}{2}} ~~~ (T<T_c)
\label{spinoregion}
\end{equation}
\noindent this region is called the spinodal region and corresponds to 
thermodynamically unstable states. The lines   $ M_s(T) $ vs. $ T $ and
$ M_0(T) $ vs. $ T $ [see eq.(\ref{MofT})] are known 
as the classical spinodal and coexistence  lines respectively. Fig. 2 displays the classical spinodal and coexistence curves for the potential $V[M]$ in (\ref{freenergy}).

 The states
between the spinodal and coexistence lines are metastable (in mean-field 
theory). As the system is cooled below $T_c$ into the unstable region inside
the spinodal, the {\em equilibrium} state of the system is a coexistence of
phases separated by domains and the concentration of phases is determined by the Maxwell 
construction and the lever rule. 




{\bf Question:} How to describe the {\em dynamics} of the phase transition and
the process of phase separation?

{\bf Answer:} A phenomenological but experimentally succesful description,
Time Dependent Ginzburg-Landau theory (TDGL) where the basic ingredient is
 Langevin dynamics\cite{bray}-\cite{marco}
\begin{equation}
\frac{\partial M(\vec r,t)}{\partial t} = -\Gamma[\vec r,M]  \; 
\frac{\delta F[M]}{\delta M(\vec r,t)} + \eta(\vec r,t) \label{langevin}
\end{equation}
with $ \eta(\vec r,t) $
a stochastic noise term, which is typically assumed to be white (uncorrelated)
and Gaussian and obeying the fluctuation-dissipation theorem:
\begin{equation}
\langle \eta(\vec r,t) \eta(\vec r',t') \rangle = 2\,T\, \Gamma(\vec
r) \, \delta^3(\vec r - {\vec r}^{\prime})\delta(t-t') ~~ ; ~~ \langle
\eta(\vec 
r,t) \rangle =0 \label{noise}
\end{equation} 
 \noindent the averages $\langle \cdots \rangle$ are over the Gaussian 
distribution function of the noise. There are two important cases to 
distinguish: {\bf NCOP:} Non-conserved order parameter, with $\Gamma=\Gamma_0$
a constant independent of space, time and order parameter, and which can
be absorbed  in a rescaling of  time. {\bf COP:} Conserved
order parameter with 
$$
\Gamma[\vec r] = - \Gamma_0 \; \nabla^2_{\vec r}
$$ 
where $\Gamma_0$ could depend on the order parameter, but  here we
will restrict the discussion to the case where it is a constant.
 In this latter case the average over the noise of the Langevin equation
can be written as a conservation law 
\begin{eqnarray}
\frac{\partial M}{\partial t} & = &  -\nabla\cdot J + \eta \; \Rightarrow \;
\frac{\partial}{\partial t} \langle \int d^3\; \vec r M(\vec r,t) \rangle =0 
\nonumber \\
\vec J & = &  \vec{\nabla}_{\vec r}
\left[-\Gamma_0\frac{\delta F[M]}{\delta M}
\right] \equiv \vec{\nabla}_{\vec r}\mu \label{equcop}
\end{eqnarray} 
\noindent where $\mu$ is recognized as the chemical potential. Examples
of the NCOP are the magnetization in ferromagnets, the gap in superconductors
and the condensate density in superfluids (the total particle number
is conserved but not the condensate fraction), of the COP: the concentration
difference in binary fluids or alloys. For a quench from $T>T_c$ deep into the
low temperature phase $T\rightarrow 0$ the thermal fluctuations are suppressed
after the quench and the noise term is irrelevant. In this situation of
experimental relevance of a deep quench the dynamics is now described by
a deterministic equation of motion, 

 for {\bf  NCOP}:
\begin{equation}
\frac{\partial M}{\partial t} = -\Gamma_0 \; \frac{\delta F[M]}{\delta
M}\label{NCOPeqn} 
\end{equation}

\noindent for {\bf  COP}:
\begin{equation}
\frac{\partial M}{\partial t} = \nabla^2\left[
\Gamma_0 \frac{\delta F[M]}{\delta M}\right] \label{cahnhil}
\end{equation}
\noindent which is known as the Cahn-Hilliard equation\cite{bray,langer}. In both cases
the equations of motion are purely diffusive
\begin{equation}
\frac{dF}{dt} = \int d^3 r \; \frac{\delta F[M]}{\delta M(\vec r,t)}  \;
\frac{\partial M(\vec r,t)}{\partial t} = -\Gamma_0
 \left\{ \begin{array}{cc}
& \int d^3 r \left(\frac{\delta F}{\delta M}\right)^2 
 ~\mbox{NCOP}  \\
& \int d^3 r \left(\vec{\nabla} \frac{\delta F}{\delta M}\right)^2 
 ~\mbox{COP} 
\end{array} 
\right. \label{diffu}
\end{equation}
\noindent and in both cases $\frac{dF}{dt}<0$. Thus, the energy is always
diminishing and there is no possibility of increasing the free energy. Thus
overbarrier thermal activation cannot be described in the absence of thermal
noise, which is clear since thermal activation is mediated by large thermal
fluctuations.  The fact that this phenomenological description is purely
dissipative with an ever diminishing free energy is one of the fundamental
differences with the quantum field theory description studied in the next
sections. 

\subsection{Critical slowing down in NCOP:}
\noindent Critical slowing down of long-wavelength fluctuations is
built in the TDGL description. Consider 
the case of NCOP and linearize the TDGL equation above the critical
temperature for small amplitude 
fluctuations near $M = 0$. Neglecting the noise term for the moment
and taking the Fourier transform of the small 
amplitude fluctuations we find
\begin{equation}
\frac{d m_k(t)}{dt} \approx -\Gamma_0\left[k^2+ r_0(T-T_c)\right]m_k(t) 
\end{equation}
showing that long-wavelength small amplitud fluctuations relax to
equilibrium $m_k =0$ on a time scale given by 
\begin{equation}
\tau_k \propto \left[k^2+ r_0(T-T_c)\right]^{-1} \label{relax}
\end{equation}
As $T \rightarrow T_c^+$ the long-wavelength modes are critically
slowed down and relax to equilibrium on very long 
time scales. Therefore a TDGL description  leads to the conclusion
that if the cooling rate is finite, 
the long-wavelength modes will fall out of LTE and become quenched. As
the temperature falls below the critical, 
these modes will become unstable and will grow exponentially. 

\subsection{Linear instability analysis:}

Let us consider now the situation for $T<<T_c$ and neglect the thermal
noise. The early time evolution after the 
quench is obtained by  linearizing the TDGL equation around a homogeneous
mean field  solution $M_o(t)$. Writing
\begin{equation}
M(\vec r,t) = M_o(t) + \frac{1}{\sqrt{\Omega}} \sum_{\vec k \neq 0}
m_k(t) \; e^{i \vec k\cdot \vec r}
\end{equation} 
where $\Omega$ is the volume of the system, and considering only the 
linear term in the fluctuations $m_k(t)$ the linearized dynamics is the following: 
{\bf COP:} for $M_o(t)$ the conservation gives
$$
\frac{d M_o(t)}{dt} =0
$$
since $M_o$ is the volume integral of the order parameter [see
eq.(\ref{equcop})]   and for the fluctuations we obtain
\begin{equation}
\frac{d m_k(t)}{dt} =   \omega(k) \; m_k(t) ~~;  ~~ \omega(k)  =
-\Gamma_0 \; k^2\left[k^2 + \left. 
\frac{\partial^2 V[M]}{\partial M^2}\right|_{M_o} \right] \label{freqcop}
\end{equation}
In the spinodal region 
$\left.\frac{\partial^2 V[M]}{\partial M^2}\right|_{M_o} <0$
there is a band of  unstable wave vectors $k^2 < 
\left| \left. \frac{\partial^2 V[M]}{\partial M^2}\right|_{M_o}\right|
$ for which 
the frequencies are positive and the fluctuations away from the mean
field grow exponentially. 

{\bf NCOP:} separate the $\vec k\neq 0$ from the $\vec k=0$ in the
linearized equation of motion:
\begin{eqnarray}
\frac{d M_o(t)}{dt} & = &  -\Gamma_0 \left.\frac{d V[M]}{dM}\right|_{M_o}
\label{zeromodelang} \\
\frac{d m_k(t)}{dt} & = & -\Gamma_0 \left[\frac{\delta F[M]}{\delta M}
\right]_{M_o(t)} m_k(t)~  = -\Gamma_0 \left[k^2 + \left.
\frac{\partial^2 V[M]}{\partial M^2}\right|_{M_o} \right] \label{kmodeslang}
\end{eqnarray}
whereas the first equation (\ref{zeromodelang}) determines that $
M_o(t) $
rolls down the potential hill towards the equilibrium solution,
the second equation also displays the linear instabilities for the
same band of wave vectors as in the COP in the spinodal region $ |M_o(t)|\leq
M_s(T) $ [see eq. (\ref{spinoregion})] for which the fluctuations grow
exponentially in time. Thus in the linearized approximation for 
both NCOP and the COP the spinodal 
instabilities are manifest as exponentially growing fluctuations. These
instabilities are the hallmark of the process of phase separation and 
are the early time indications of the formation and growth of correlated
regions which will be understood in an exactly solvable example below.

\subsection{The scaling hypothesis: dynamical length scales for ordering}
The process of ordering is described by the system developing ordered regions
or domains that are separated by walls or other type of defects. 
The experimental probe to study the domain structure and the emergence of
long range correlations is the equal time pair correlation function
\begin{equation}
C(\vec r,t) = \langle M(\vec r,t) M (\vec 0,t) \rangle
\end{equation} 
where $\langle \cdots \rangle$ stands for the statistical ensemble average
in the initial state (or average over the noise in the initial state before
the quench) and will become clear(er) below.  It is convenient to expand
the order parameter in Fourier components
$$
M(\vec r,t) = \frac{1}{\sqrt{\Omega}} \sum_{\vec k} m_k(t)  \; 
e^{i\vec k\cdot\vec x}
$$
and to consider the spatial Fourier transform of the pair correlation function
\begin{equation}
S(\vec k,t) = \langle m_{\vec k}(t) m_{-\vec k}(t) \rangle \label{strucfac}
\end{equation}
known as the {\bf structure factor} or power spectrum which is experimentally
measured by neutron (in ferromagnets) or light scattering (in binary fluids)\cite{goldburg}.
The scaling hypothesis introduces a dynamical length scale $L(t)$ that
describes the typical scale of a correlated region and proposes that
\begin{equation}
C(\vec r,t) = f\left(\frac{|\vec r|}{L(t)}\right) \Rightarrow S(\vec k,t) =
L^d(t) \;  g(kL(t)) \label{scaling}
\end{equation}
\noindent where $d$ is the spatial dimensionality and 
$f$ and $g$ are scaling functions. Ultimately scaling is confirmed by
experiments and numerical simulations and theoretically it emerges from
a renormalization group approach to dynamical critical phenomena which
provides a calculational framework to extract the scaling functions and
the deviations from scaling behavior\cite{bray}. This scaling hypothesis describes
the process of phase ordering as the formation of ordered `domains' or
correlated regions of typical spatial size $L(t)$. For NCOP typical growth
laws are $L(t) \approx t^{1/2}$ (with some systems showing weak logarithmic
corrections) and $L(t) \approx t^{1/3}$ for scalar and $\approx t^{1/4}$
for vector order parameter in the COP case\cite{bray,mazenko,marco}. 

\subsection{An exactly solvable (and relevant) example: the Large $ N $ limit}

We consider the case where the order parameter has $N$-components and transforms
as a vector under rotations in an N-dimensional Euclidean space, i.e.
$\vec M(\vec r,t)= (M_1(\vec r,t), M_2(\vec r,t), \cdots, M_N(\vec r,t))$.
For $N=1$ an example is the Ising model, for $N=2$ superfluids or 
superconductors (where the components are the real and imaginary part of
the condensate fraction or the complex gap respectively), $N=3$ is the spin
one Heisenberg antiferromagnet, etc. For $N=1$ the  topological defects
 are domain walls (topological in one spatial dimension), for $N=2$ they
are vortices in $d=2$ and vortex lines in $d=3$, for $N=d=3$ the topological
defects are monopoles or skyrmions which are possible excitations in
Quantum Hall systems and also appear in nematic liquid
crystals\cite{kibble2}. For  
$N\rightarrow \infty$ and fixed $d$ no topological defects exist. However
the exact solution of the large $ N $ model gives insight and is in fairly
good agreement with growth laws for fixed $ N $ systems which had been studied
experimentally and numerically\cite{bray,marco}.  In cosmological
space-times  it has
been implemented to study the collapse of texture-like
configurations\cite{durrer,turok,filipe} (see 
later). In quantum field
theory the non-equilibrium dynamics of  
phase transitions  has been studied  in Minkowsky and cosmological
space-times\cite{boydcc,FRW,nuestros,noscorre,losalamos,new}. The large $ N $ limit is an exactly solvable limit that serves
as a testing 
ground for establishing the fundamental concepts and that can be systematically
improved in a consistent $1/N$ expansion. It provides a consistent
formulation which is {\em non-perturbative}, 
renormalizable and numerically implementable and has recently been
invoked in novel studies of non-equilibrium dynamics in quantum spin glasses
and disordered systems\cite{cugliandolo}.  

 The exact solution for the dynamics in the large $ N $ limit, being
available both in 
the condensed matter TDGL description of phase ordering kinetics and
in Quantum Field 
Theory in Minkowsky and Cosmological space times, allow us to compare
{\em directly} the physics of phase ordering in these situations. Thus
we begin by implementing this 
scheme in the NCOP case for the TDGL description. 
 
What is the $ \langle \cdots \rangle $ in the equations of the previous
section?: consider that 
{\em before} the quench the system in in equilibrium in the {\em disordered}
phase at $ T>>T_c $ and with a very short correlation length ($ \xi(T)
\approx 1/T $). The  
ensemble average in this initial state is therefore 
\begin{eqnarray}
\langle M^i(\vec r,0) M^j({\vec r}',0) \rangle & = & \Delta  \;
\delta^{ij} \delta^3(\vec r-{\vec r}') \nonumber \\ 
\langle M^i(\vec r,0)  \rangle & = & 0 \label{inicor}
\end{eqnarray} 
\noindent where $\Delta$ specifies the initial correlation. Now
consider a critical quench where the system 
is rapidly cooled through the phase transition to almost zero
temperature but in the {\em absence} of explicit symmetry 
breaking fields (for example a magnetic field). The average of the
order parameter will remain zero through the 
process of spinodal decomposition and phase ordering. During the
initial stages, linear instabilities will grow 
exponentially with $m^i_k(t) \approx m^i_k(0) \; e^{\omega(k) t}~~; ~~
\omega(k)= k^2-r(0)$ for $k^2 < r(0)$ and 
at early times
\begin{equation}
\langle m^i_{\vec k}(t) m^j_{-\vec k}(t) \rangle \approx \Delta \;
e^{2\omega(k)t} 
\end{equation}  
hence fluctuations begin to grow exponentially and eventually will
sample the broken symmetry states and the exponential 
growth must shut-off.
The large $ N $ limit is implemented by writing the potential term in
the free energy as 
\begin{equation}
V[\vec M] = -\frac{r(T)}{2} \; \vec M^2 + \frac{\lambda}{4N} (\vec M^2)^2
~; ~~ \vec M^2= \vec M \cdot \vec M 
\label{largenpot}
\end{equation}
where $\lambda$ is kept finite in the large $ N $ limit. We will focus
on the NCOP case with a quench to zero temperature and rescale the
order parameter, time and space 
as 
\begin{equation}
\vec{M} = \sqrt{\frac{r(0)}{\lambda}} \; \vec \eta ~~; ~~ r(0) \;\Gamma_0 \; t
= \tau ~~; ~~ \sqrt{r(0)} \;\vec x = z 
\label{rescalevars}
\end{equation}
after which the evolution equation for the NCOP case becomes
\begin{equation}
\frac{\partial \vec \eta}{\partial \tau} = \nabla^2 \vec \eta + \left(1-\frac{{\vec{\eta}}^2}{N}\right)~\vec \eta
\end{equation}
where derivatives are now with respect to the rescaled variables. The large $ N $ limit is solved by implementing
a Hartree-like factorization\cite{bray}
\begin{equation}
\vec \eta^2 \rightarrow \langle \vec \eta^2 \rangle =  N  \langle \eta^2_i \rangle ~~ \mbox{no sum over i}
\label{largeNfactori}
\end{equation}
Then for each component the NCOP equation becomes
\begin{eqnarray}
\frac{\partial  \eta_i}{\partial \tau} & = & \left[\nabla^2
+M^2(t)\right] \eta_i \label{effeqn} \\ 
M^2(t) & = & 1-\langle \eta^2_i \rangle  \label{selfconsncop}
\end{eqnarray}
the eq.(\ref{selfconsncop}) is a {\em self-consistent} condition that
must be solved simultaneously with the 
equation of motion for the components. Thus the large $ N $
approximation linearizes the problem at the expense of 
a self-consistent condition. The solution for each component is obviously
\begin{equation}
\eta_{i,\vec k}(\tau) = \eta_{i,\vec k}(0) \;e^{-k^2 \tau + b(\tau)}
~; ~~ b(\tau) = \int^{\tau}_0 M^2(\tau')d\tau' 
\end{equation}
Consider for a moment that the $\vec k=0$ mode is slightly displaced
at the initial time, then it will roll 
down the potential hill to a final equilibrium position for which
$ M^2(\infty) \;\eta_i(\infty) =0 $ (so the 
time derivative vanishes in equilibrium). If $ \eta_i(\infty)\neq 0 $ is
a broken symmetry minimum of the free energy, 
then $ M^2(\tau) \rightarrow  0$ when $ \tau \rightarrow \infty $. This is
the statement of Goldstone's theorem that 
guarantees that the perpendicular fluctuations are soft modes. This
asymptotic limit allows the solution of the 
self-consistent condition  
\begin{equation}
M^2(\tau) = 1-\langle \eta^2_i(\tau) \rangle = 1- {\Delta} \; e^{2
b(\tau)} \int \frac{d^d k}{(2\pi)^d} \;  e^{-k^2 \tau} = 
1- {\Delta} \;  e^{2b(\tau)} \; (8\pi t)^{-\frac{d}{2}}
\end{equation}
The vanishing of the right hand side in the asymptotic time regime
leads to the self-consistent solution 
\begin{equation}
b(\tau) \rightarrow \frac{d}{4} \ln\left[\frac{\tau}{\tau_0}\right] ~~ \Rightarrow M^2(\tau) \rightarrow \frac{d}{4\tau}
\end{equation}
where $\tau_0$ is a constant related to $\Delta$. This self-consistent solution results in the following
 asymptotic behavior
\begin{equation}
\eta_{i,\vec k}(\tau) \rightarrow \eta_{i,\vec k}(0)\left(\frac{\tau}{\tau_0}\right)^{\frac{d}{4}} e^{-k^2\tau}
\end{equation}
Introducing the {\em dynamical length scale} $L(\tau)= \tau^{\frac{1}{2}}$ it is straightforward to find
the structure factor and the pair correlation function
\begin{eqnarray}
&&S(\vec k,t) \propto L^d(t) \; e^{-2(kL(t))^2} \label{struclargen} \\
&&C(\vec r, t) \propto e^{-\frac{r^2}{8L^2(t)}} ~~;~~ L(t) =
t^{\frac{1}{2}} \label{paircorrlargen} 
\end{eqnarray}
This behavior {\em should not} be interpreted as diffusion, because of the
$L^d(t)$ in eqn. (\ref{struclargen}) which is a result of the self-consistent condition.

\underline{\bf Important Features:}
\begin{itemize}
\item{The `effective squared mass' $M^2(t)\stackrel{t\rightarrow
\infty}{\rightarrow} 0$: asymptotically there 
are massless  excitations identified as Goldstone bosons.}

\item{Since $M^2(t)\rightarrow 0$ asymptotically, the self-consistent
condition results in that $\langle \vec M^2 \rangle 
\rightarrow N r(0)/\sqrt{\lambda}$, i.e. the fluctuations sample the
broken symmetry states, which are equilibrium 
minima of the free energy. These fluctuations begin to grow
exponentially at early times due to spinodal instabilities.}  

\item{A dynamical correlation length emerges $L(t)= t^{1/2}$ which
determines the size of the correlated regions or 
`domains'. A scaling solution emerges asymptotically with the
natural scale determined by the size of the ordered 
regions. These regions grow with this law until they become
macroscopically large.  
Although this a result obtained in the large $ N $ limit, similar
growth laws had been found for NCOP both 
analytically and numerically for $N=1$ etc.\cite{bray}} 

\item{{\bf Coarsening:} The expression for the structure factor (\ref{struclargen}) shows that at large times
only the very small wavevectors contribute to $S(\vec k,t)$, however the self-consistency condition forces the
$\int k^{d-1} \; dk \; S(k,t) \rightarrow \mbox{constant}$ thus
asymptotically $k^{d-1}~S(k,t)$ is 
peaked at wavevectors $k \approx L^{-1}(t)$ with an amplitude $L^d(t)$ thus becoming a {\em delta
function} $S(\vec k,t) \stackrel{t\rightarrow \infty}{\rightarrow} \delta^d(\vec k)$. The position of the peak
in $S(\vec k,t)$ moving towards longer wavelength is the phenomenon of coarsening and is observed via light
scattering. At long times a zero momentum condensate is formed\cite{marco} and a Bragg peak develops at zero momentum, this
condensate however grows as a power of time and only becomes macroscopic at asymptotically large times. Coarsening
is one of the experimental hallmarks of the process of phase ordering, revealed for
example in light scattering\cite{goldburg} and is found numerically in many systems\cite{bray}. Thus
the large $ N $ limit, although not being able to describe topological defects offers a very good description of
the ordering dynamics.} 
\end{itemize}

\section{Phase ordering in Quantum Field Theory I: Minkowski space-time}
\subsection{A quench in Q.F.T.}
Although the phenomenological Time Dependent Landau Ginzburg theory is a
succesful  description of phase ordering kinetics in condensed
matter systems, there is no first principle derivation from a microscopic theory of these equations of motion. Whereas microscopic descriptions either based on classical or quantum Hamiltonians lead to time reversal invariant equations of motion, the TDGL equations are first order in the time derivative and therefore purely dissipative. 

A first principles, microscopic description of a quantum theory must begin
with the Heisenberg equations of motion for operators or the Schroedinger
or quantum Liouville equations for the quantum states or density matrix that describes the system. In this section we provide an introduction to the
treatment of strongly out of equilibrium situations, in particular that
of a ``quench'' in a quantum field theory system. 

This is the situation studied in\cite{broken} for the dynamics of formation and evolution of disoriented chiral condensates during the chiral phase transition.  

The dynamics is completely determined by the microscopic field theoretical
Hamiltonian. For a simple scalar theory
the Hamiltonian operator is given by
\begin{equation}
\hat{H} = \int d^3x \left\{ \frac{1}{2}{\Pi}^2(\vec x,t)+
\frac{1}{2}[\vec{\nabla}\Phi(\vec x,t)]^2 + V[\Phi(\vec x,t)]\right\}
\label{qftham} 
\end{equation} 
where $\Phi$ is the quantum mechanical field and $\Pi$ its canonical
momentum. We want 
to describe a quenched scenario where the initial state of the system
for $t<0$ is the ground state (or density matrix,  
see later) of a Hamiltonian for which the potential is convex for all
values of the field, for example that of an 
harmonic oscillator, in which case the wave function(al) $\Psi[\Phi]$
is a Gaussian centered at the origin. At $t=0$ the 
potential is changed so that for $t>0$ it allows for broken symmetry
states. This can be achieved for example 
by the following form
\begin{eqnarray}
V[\Phi] & = & \frac{1}{2} m^2(t) \Phi^2 + \frac{\lambda}{4}\Phi^4 \label{qftpotential} \\
m^2(t) & = & \left\{ \begin{array}{cc}
+m^2_0>0 & ~\mbox{for} ~ t<0 \\
-m^2_0<0 &  ~\mbox{for} ~t>0
\end{array} 
\right. \label{massat}
\end{eqnarray}
thus the potential in Fig. 1 changes {\em suddenly} from $T>T_c$ to
$T<T_c$. 
Although in Minkowski space-time this is an {\em ad-hoc} choice of a
time dependent potential that 
mimics the quench\cite{bowick}, we will see in the next section that
in a cosmological setting the mass term naturally 
depends on time through the temperature dependence and that it changes
sign below the critical temperature as the 
Universe cools off. Most of the results obtained in Minkowski
space-time will translate onto analogous results in 
a Friedmann-Robertson-Walker cosmology. Unlike the phenomenological
(but succesful) description of the dynamics in 
condensed matter systems, in a microscopic quantum theory the dynamics
is completely determined by the Schr\"odinger 
equation for the time evolution of the wave function or alternatively
the Liouville equation for the evolution of 
the density matrix in the case of mixed states. We will cast our study
in terms of a density matrix in general, such 
a density matrix could describe pure or mixed  states and  obeys the
quantum Liouville equation 
\begin{equation}
i \frac{\partial \hat{\rho}(t)}{\partial t} =
\left[\hat{H}(t),\hat{\rho}(t)\right] \label{liouville}  
\end{equation} 
{\bf Question:} How does the wave function(al) or the density matrix
evolve after a quench?  
\subsection{A simple quantum mechanical picture:}
In order to gain insight into the above
question, let us consider a simple case of one quantum mechanical
degree of freedom $q$ and the quench is 
described in terms of an harmonic oscillator with a time dependent frequency 
$\omega^2(t)= -\epsilon(t) \; \omega^2_0 ~;~\omega^2_0>0$
with $ \epsilon(t) $ the sign function, so that 
$\omega^2(t<0)>0 ~;~\omega^2(t>0)<0$. Furthermore let us focus 
on the evolution of  a pure state (the density matrix is simple the product
of the wave function and its complex conjugate). Consider that at $t<0$ the
wave function corresponds to the ground state of the (upright) harmonic
oscillator.  For $t>0$ the wave function
obeys
\begin{equation}
i\frac{\partial \Psi[q,t]}{\partial t} = \left[-\frac{1}{2}
\frac{d^2}{dq^2}-\frac{1}{2}\omega^2_0\; q^2 \right]\Psi[q,t] 
\end{equation}  
Since the initial wave function is a gaussian and under time evolution with a quadratic Hamiltonian Gaussians remain
Gaussians, the solution of this Schr\"odinger equation is given by
\begin{eqnarray}
\Psi[q,t] & = & N(t) \;  e^{-\frac{A(t)}{2}q^2} \label{qmwf} \\
 \frac{d ln N(t) }{dt} & = & -\frac{i}{2} A(t)  \label{qmunit} \\
i \frac{dA}{dt} & = & A^2 + \omega^2_0 \label{qmkern} \nonumber
\end{eqnarray}

Separating the real and imaginary parts of $A(t)$ it is straightforward to
find that $ |N(t)|^4/\mbox{Re}[A(t)] $ is constant,  a consequence
of unitary time evolution. Eq.(\ref{qmkern}) can be cast in a more
familiar form by a simple substitution
\begin{equation}
A(t) = -i\frac{\dot{\phi}(t)}{\phi(t)} \Rightarrow \ddot{\phi}(t)-\omega^2_0
 \; \phi(t)=0
\end{equation}
where the equation for $\phi$ was obtained by inserting the above expression
for $A(t)$ in (\ref{qmkern}). The solution is $\phi(t) = a \;  e^{\omega_0~t}+
b \;  e^{-\omega_0~t}$ featuring exponential growth. This is the
quantum mechanical 
analog of the spinodal instabilities described in the previous section. 
The equal time two-point function is given by
\begin{equation}
\langle q^2 \rangle(t) = A^{-1}_R(t) = |\phi(t)|^2 \approx
e^{2\omega_0 \, t}
\end{equation}
The width of the Gaussian state increases in time (while the amplitude
decreases to maintain a constant norm) and the quantum fluctuations grow
exponentially. As the Gaussian wave function spreads out the probability for
finding configurations with large amplitude of the coordinates increases.
These is the quantum mechanical translation of the linear spinodal
instabilities. When the non-linear contributions to the quantum mechanical
potential are included the single particle quantum mechanical
wave function will simply develop two peaks and eventually re-collapse by
focusing near the origin undergoing oscillatory motion between `collapses'
and `revivals'. In the case of a full quantum field theory
there are infinitely many degrees of freedom and the energy is transferred
between many modes. This simple quantum mechanical example paves
the way for understanding in a simple manner the main features of a quench in the
large $ N $ limit in quantum field theory, to which we now turn our attention. 
\subsection{Back to the original question: Large $ N $ in Q.F.T.}
We now consider the large $ N $ limit of a full Q.F.T. in which
\begin{equation}
\vec{\Phi}(\vec x,t)= \left(\Phi_1(\vec x,t), \Phi_2(\vec x,t), \cdots,
\Phi_N(\vec x,t) \right) 
\end{equation}
and similarly for the canonical momenta $\vec{\Pi}$. The Hamiltonian
operator is of the form (\ref{qftham})
with
\begin{equation}
V[\vec \Phi]  =  \frac{1}{2} m^2(t) \;  \vec{\Phi}\cdot\vec{\Phi}
 + \frac{\lambda}{8N}[\vec{\Phi}\cdot\vec{\Phi}]^2 \label{qftpotlargen} 
\end{equation}
with $m^2(t)$ given by (\ref{massat}). Let us focus on the case in which 
the initial state pure and symmetric, i.e. $\langle \Phi \rangle =0$, with $<\cdots>$ being
the expectation value in this initial state. The more complicated case
of a mixed state, described 
by a density matrix is studied in detail
in\cite{nuestros,noscorre,losalamos} and the main features are the
same as those revealed 
by the simpler scenario of a pure state.  The large $ N $ limit is implemented in a similar manner as in the
TDGL example, via a Hartree like factorization
\begin{equation}
(\vec{\Phi}\cdot \vec{\Phi})^2 \rightarrow 2\, \langle \vec{\Phi}\cdot
\vec{\Phi}\rangle \; \vec{\Phi}\cdot \vec{\Phi} 
\label{largenqft}
\end{equation}  
where the expectation value is in the time evolved quantum state (in the Schr\"odinger picture) or in the initial
state of the Heisenberg operators (in the Heisenberg picture). Via this factorization the Hamiltonian becomes quadratic at
the expense of a self-consistent condition as it will be seen below. It is convenient to introduce the spatial Fourier transform of the fields as
\begin{equation}
\vec{\Phi}(\vec x,t) = \frac{1}{\sqrt{\Omega}} \sum_{\vec k}
\vec{\Phi}_{\vec k}(t) \;  e^{i \vec k \cdot \vec x} 
\end{equation}
with $\Omega$ the spatial volume, and a similar expansion for the
canonical momentum $\Pi(\vec x,t)$. The Hamiltonian becomes
\begin{eqnarray}
H & = & \sum_{\vec k} \left\{ \frac{1}{2} \vec{\Pi}_{\vec k}\cdot \vec{\Pi}_{-\vec k}+ 
\frac{1}{2} W^2_k(t)\; \vec{\Phi}_{\vec k}\cdot \vec{\Phi}_{-\vec k}
\right\} \label{hamqftlargen} \\ 
W^2_k(t) & = & m^2(t)+k^2+\frac{\lambda}{2N}\int \frac{d^3k}{(2\pi)^3}\; 
 \langle \vec{\Phi}_{\vec k} \cdot \vec{\Phi}_{-\vec k}  \rangle(t) 
\label{timefreqs} 
\end{eqnarray}
The problem now
has decoupled in a set of infinitely many harmonic oscillators, that
are only coupled through the self-consistent 
condition in the frequencies (\ref{timefreqs}). To induce a quench,
the time dependent mass term has the form proposed in eq. (\ref{massat}).

Just as in the simple quantum mechanical case, we consider the initial
state to be a Gaussian centered at the origin in field space, 
which is the ground state of the (upright) harmonic oscillators for
$t<0$. Since a Gaussian is always a Gaussian under 
time evolution with a quadratic Hamiltonian, we propose the wave
function(al) that describes the (pure) quantum mechanical 
state to be given by
\begin{equation}
\Psi[\vec{\Phi},t] = \Pi_k\left\{ N_k(t) \; e^{-\frac{A_k(t)}{2}
\vec{\Phi}_{\vec k}\cdot \vec{\Phi}_{-\vec k}} \right\} 
~;~~ A_k(t=0)= W_k(t<0) \label{wavefunc}
\end{equation}
Time evolution of this wavefunction(al) is determined by the
Schr\"odinger equation: in the Schr\"odinger representation 
the canonical momentum becomes a differential (functional) operator, 
$\vec{\Pi}_{\vec k} \rightarrow -i\delta/\delta \vec{\Phi}_{-\vec k}$
and the Schr\"odinger equation becomes a functional 
differential equation. Comparing the powers of $\Phi_{\vec k}$ in this
differential equation, one obtains the following evolution 
equations for $N_k(t)$ and $A_k(t)$ 
\begin{eqnarray}
\frac{d}{dt}\ln N_k(t) & = & -\frac{i}{2} A_k(t) 
\label{normqft}\\
i \frac{dA_k(t)}{dt} & = & A^2_k(t)- W^2_k(t) \label{kernqft} 
\end{eqnarray}
As in the single particle case, the constancy of
$ |N_k(t)|^4/\mbox{Re}[A_k(t)] $ is a consequence of unitary time  
evolution. The non-linear equation for the kernel $ A_k(t) $ can be
simplified just as in the single particle case by writing
\begin{equation}
A_k(t) = -i \frac{\dot{\phi}_k(t)}{\phi_k(t)} \Rightarrow 
\ddot{\phi}_k(t)+W^2_k(t)\;  \phi_k(t)=0 \label{modesqft}
\end{equation}
and taking the expectation value of $\Phi^2 $ in this state we obtain
\begin{equation}
\langle \vec{\Phi}_{\vec k} \cdot \vec{\Phi}_{-\vec k}  \rangle(t)= N\;
|\phi_k(t)|^2  
\end{equation}
Hence we find a self-consistent condition much like the one obtained
 in the large $ N $ limit for TDGL. The equations for the mode
functions and the self-consistent condition for $t>0$ are therefore given
by 
\begin{eqnarray}
&& \ddot{\phi}_k(t)+[k^2+M^2(t)]\; \phi_k(t) =  0 \label{eqnqftlargen} \\
&&M^2(t) =  -m^2_0 + \frac{\lambda}{2} \int \frac{d^3k}{(2\pi)^3} |\phi_k(t)|^2
\label{selfconsqft} 
\end{eqnarray}
where the integral in the self-consistent term in (\ref{selfconsqft}) is 
simply $\langle \Phi^2_i \rangle$.
There are two fundamental {\em differences} between the quantum dynamics
determined by the equations of motion and the classical dissipative
dynamics of the TDGL phenomenological description given in sec. II:
\begin{itemize}
\item{The equations of motion and the self-consistency condition equations
(\ref{eqnqftlargen})-(\ref{selfconsqft}) lead immediately to the conservation
of energy\cite{FRW,nuestros}.}
\item{The evolution equations are {\em time reversal invariant}.}
\end{itemize}
These properties must be contrasted to the purely dissipative evolution
dictated by the TDGL equations as is clear from eq. (\ref{diffu}).
 Consider a very weakly coupled theory
$\lambda <<1$ and very early times, then the self-consistent term can
be neglected and we see that for $k^2 <m^2_0$ the modes grow exponentially.
This instability again is the manifestation of spinodal 
growth\cite{erickwu,boyvega,boylee,nuestros,noscorre}. Since the
mode functions grow exponentially, fairly soon, at a time scale $t_s \approx
m^{-1}_0 \ln(1/\lambda)$ the self-consistent term begins to cancel the
negative mass squared and  $M^2(t)$ becomes
smaller. We find numerically that this effective mass vanishes asymptotically,
as shown in Fig. 3.

\subsection{Emergence of condensates and classicality:}
The physical mechanism here is similar to that in the classical TDGL,
but  in terms of quantum  
fluctuations. The quantum fluctuations with wave vectors inside the
spinodally unstable band grow 
exponentially, these make the $\langle \Phi^2 \rangle$ self-consistent
field to grow non-perturbatively  
large until when $\langle \Phi^2 \rangle \approx m^2_0/\lambda$ when
the self-consistent (mean) field  
begins to be of the same order as $m^2_0$ (the tree level mass
term). At this point the {\em quantum} 
fluctuations become non-perturbatively large and sample field
configurations near the  equilibrium minima of the 
potential. The spinodal instabilities are shutting off since the effective squared mass $M^2(t)$ is vanishing.

When $M^2(t)$ vanishes, the equations for the mode functions become those
of a free massless field, with solutions of the form $\phi_k(t) = A_k
\; e^{ikt}+ B_k \; e^{-ikt}$, whereas for the $k=0$ mode the solution
must be of the form 
$\phi_0(t) = a+bt$ with $a ; b \neq 0$ since the Wronskian of the mode
function and its complex conjugate is 
a constant. This in turn determines that the low $k$ (long wavelength)
behavior of the mode functions is given 
by
\begin{equation}
\phi_k(t) = a \cos kt + b \; \frac{\sin kt}{k} \label{specqft}
\end{equation}

This behavior at long wavelength has a remarkable consequence: at very
long time the power spectrum 
$ |\phi_k(t)|^2 $, which is the equivalent of $S(k,t)$ for TDGL (see
eq. (\ref{strucfac})) is dominated by the small $k$-region, in
particular $ k<<1/t $, with an amplitude that 
grows quadratically with time. Then the structure factor $ S(\vec k,t)
= |\phi_k(t)|^2 $ features a 
peak that moves towards longer wavelengths at longer times and whose
amplitude grows with time in such a  
way that asymptotically $ \int^{\infty}_0 k^2 S(\vec k,t) dk / 2\pi^2 \rightarrow
m^2_0/\lambda $ and the integral is dominated 
by a very small region in $k$ that gets narrower at longer times. This
is the equivalent of {\em coarsening} 
in the TDGL solution in the large $ N $ limit, where the asymptotic
time regime was dominated by the formation of 
a long-wavelength condensate. Fig. 4  shows the power spectrum at
two (large) times displaying clearly the 
phenomenon of coarsening and the formation of a non-perturbative condensate. 

The pair correlation function can now be calculated using this power
spectrum\cite{noscorre} 
\begin{equation}
C(\vec r,t) = \frac{1}{2 \, \pi^2 r} \int_0^{\infty} k \sin kr \; |\phi^2_k(t)|
\; dk \; . 
\end{equation}
At long times and distances the integral is dominated by the very long wavelength modes,
in particular by the term $\propto \sin[kt]/k$ of $\phi_k(t)$, hence the integral can be done analytically and
we find
\begin{equation}
C(\vec r,t) = \frac{A}{r}\; \Theta(2t-r) \label{correqft}
\end{equation}
with $ A $ a constant. This is a remarkable result: the correlation
falls off as $1/r$ inside domains that 
grow at the speed of light.  This correlation function is shown in Fig. 5  at several different (large) times. This
 correlation function is of the {\em scaling form}: introducing the
dynamical length scale $L(t) = t$ it is clear that\cite{noscorre} 
\begin{equation}
C(\vec r,t) \propto L^{-1}(t) f(r/L(t)) ~~; ~~ f(s) =
\frac{\Theta(2-s)}{s} \label{scalingqft} 
\end{equation}

We interpret these `domains' as being a non-perturbative condensate of
Goldstone bosons, 
with a non-perturbatively large number of them $\propto 1/\lambda$,
such that the mean square root fluctuation of the field samples the
(non-perturbative) 
equilibrium minima of the potential. In particular an important
conclusion of this analysis is that the long-wavelength 
modes acquire very large amplitudes, their phases vary slowly as a
function of time (for $k<<1/t$), therefore these 
fluctuations which began their evolution as being quantum mechanical,
now have become {\em classical}.




\subsection{O.K...O.K. but where are the defects?}
At this point our analysis begs this question. To understand the
answer it is convenient to back track the analysis to 
the beginning. The initial quantum state is given by a the
wave-function(al) (\ref{wavefunc}), thus the most 
probable field configurations found in this ensemble are those whose
spatial Fourier transform are given by 
\begin{equation}
|\Phi_k| \propto \frac{1}{\sqrt{W_k(t<0)}} \propto \frac{1}{\sqrt{k^2+m^2_0}}
\end{equation} 
(restoring $\hbar$ would multiply $\Phi_k$ by $\sqrt{\hbar}$). Then typical long-wavelength field configurations
that are represented in the quantum ensemble described by this initial wave-function(al) are of rather small 
amplitude. The initial correlations are also rather short ranged on scales $m^{-1}_0$. Under time evolution  the probability distribution is given by 
\begin{equation}
{\cal P}[\Phi,t] = |\Psi[\Phi,t]|^2 = \Pi_{i=1}^N\Pi_k\left\{|N_k(t)|^2 e^{-\frac{|\Phi^i_k(t)|^2}{|\phi_k(t)|^2}}\right\}
\end{equation} 
At times longer than the regime dominated by the exponential growth of the spinodally unstable modes, the power
spectrum $|\phi^2_k(t)|^2$ obtains the largest support for long wavelengths $k<<m^2_0$ and with amplitudes 
$\approx m^2_0/\lambda$. Therefore field configurations with typical spatial Fourier transform $\phi_k(t)$ are very
likely to be found in the ensemble. These field configurations are primarily made of long-wavelength modes and their
amplitudes are non-perturbatively large, of the order of the amplitude of the fields in the broken symmetry minima. 
A typical such configuration can be written as
\begin{equation}
\Phi^i(\vec x,t)_{typical} \approx \sum_k |\phi_k(t)| \cos[\vec k\cdot \vec x + \delta^i_{\vec k}]
\end{equation}
where the phases $\delta^i_{\vec k}$ are randomly distributed with a
Gaussian probability distribution since the density 
matrix is gaussian in this approximation. We note
that a particular choice of these phases leads to a realization of
a likely configuration in the ensemble that 
{\em breaks translational invariance}. In fact translations can be
absorbed by a change in the phases, thus averaging 
over these random phases restores translational invariance. Since the
quantum state (or density matrix) is translational 
invariant a particular spatial profile for a field configuration
corresponds to a particular representative of 
the ensemble. Combining all of the above results together we can
present the following consistent interpretation of the ordering
process and the formation of coherent non-perturbative structures
during the dynamics of symmetry breaking in the large $ N $
limit\cite{noscorre} : 
\begin{itemize}
\item{The early time evolution occurs via the exponential growth of
spinodally unstable long wavelength modes. This 
unstable growth leads to a rapid growth of fluctuations $\langle
\Phi^2 \rangle(t)$ which in turn increases the self-consistent
contribution and tends to cancel the negative mass squared. The
effective mass of the excitations $-m^2_0+ 
\frac{\lambda}{2N}\langle \Phi^2 \rangle(t) \rightarrow 0$ and the
asymptotic excitations are Goldstone bosons.} 

\item{At times larger than the spinodal time $t_s \approx m^{-1}_0
\ln(1/\lambda)$, the effective mass vanishes and 
the power spectrum or structure factor $S(k,t)=|\phi_k(t)|^2$ displays
the features of coarsening: a peak that moves towards 
longer wavelengths and increases in amplitude, resulting in a
long-wavelength condensate at asymptotically long times.}

\item{For large time a dynamical correlation length emerges $L(t) =t$ and at
 long distances the pair correlation function is of the scaling form
$C(\vec r,t) \propto L^{-1}(t) f(r/L(t))$. The length scale $L(t)$
determines the size of the correlated 
regions and determines that these regions grow at the speed of
light. Inside these regions there is a non-perturbative 
condensate of Goldstone bosons with a typical amplitude of the order
of the value of the homogeneous field at the equilibrium broken
symmetry minima.}

\end{itemize}
The similarity between these results and those of the more
phenomenological TDGL description in condensed matter systems 
is rather striking. The features that are determined by the structure
of the quantum field theory are\cite{noscorre}: i) the scaling
variable $ s=r/t $ with 
equal powers of distance and time is  a consequence of the Lorentz
invariance of the underlying  
theory, ii) the fact that the pair correlation function vanishes for
$ r>2t $ is manifestly a consequence of causality.  
An analysis of the correlations and defect density during the spinodal
time scale has been performed in\cite{rivers} 
and related recent studies had been performed in\cite{beilok}.

\section{Phase ordering in Quantum Field Theory II: FRW Cosmology}
\subsection{Cosmology 101 (the basics):}
On large scales $> 100 ~\mbox{Mpc}$ the Universe appears to be
homogeneous and isotropic as revealed by the isotropy 
and homogeneity of the cosmic microwave background and some of the
recent large scale surveys\cite{durrer}. The cosmological principle
leads to a simple form of the metric of space time, the
Friedmann-Robertson-Walker (FRW) metric in terms of a scale factor
that determines the Hubble flow and the 
curvature of spatial sections. Observations seem to favor a flat
Universe for which the space time metric is rather 
simple:
\begin{equation}
ds^2= dt^2 - a^2(t) \; d\vec x^2 \label{FRWmetric}
\end{equation}
the time and spatial variables $ t, \vec x $ in the above metric are
called comoving time and spatial distance 
respectively and have the interpretation of being the time and
distance measured by an observer locally at rest with 
respect to the Hubble flow. At this point we must note that {\em
physical distances} are given  
by $ \vec{l}_{phys}(t)=a(t) \; \vec x $.  An important concept is that of
causal (particle) horizons: events that cannot be connected by 
a light signal are causally disconnected. Since light travels on null
geodesics $ ds^2=0 $ the maximum {\em physical} 
distance that can be reached by a light signal at time $ t $ is given  by 
\begin{equation}
d_H(t) =a(t)\int^t_0 \frac{dt'}{a(t')} \label{horizon} 
\end{equation}
It will prove convenient to change coordinates to {\em conformal time}
by defining a conformal time variable 
\begin{equation} 
\eta = \int^t_0 \frac{dt'}{a(t')} \Rightarrow ds^2 = 
C^2(\eta) \; (d\eta^2 - d\vec x^2) ~~; ~~ C(\eta) = a(t(\eta))\label{conftime}
\end{equation}
in terms of which the causal horizon is simply given by $ d_H(\eta) =
 C(\eta) \;  \eta$ and physical distances as $ \vec{x}_{phys}=C(\eta) \;  \vec
 x $. This metric is of the same form as that of  
 Minkowski space time. 
For energies well below the Planck scale $ M_{Pl} \approx
 10^{19}\mbox{Gev} $ gravitation is well described by {\em classical} 
General Relativity and the Einstein equations:
\begin{equation}
R^{\mu \nu} - \frac{1}{2} \;  g^{\mu \nu} R = \frac{8\pi}{3M^2_{Pl}}
\;T^{\mu \nu} \label{einsteineqns} 
\end{equation}
where we have been cavalier and set $c=1$ (as well as $ \hbar=1
$). $ R^{\mu \nu} $ is the Ricci tensor,  $ R $ the Ricci  
scalar and $ T^{\mu \nu} $ the matter field energy momentum
tensor. The above equation is classical but one seeks to 
understand the dynamics of the Early Universe in terms of a {\em
quantum field theory} that describes particle physics, thus the
question: what is exactly 
the energy momentum tensor?, in Einstein's equations it is a classical
object, but in QFT it is an operator. The answer to 
this question is: gravity is classical, fields are quantum mechanical, but 
$ T^{\mu \nu} \rightarrow \langle T^{\mu \nu} \rangle $, i.e. it is the
expectation value of a {\em quantum mechanical operator in a quantum
mechanical state}. This quantum mechanical state, either pure or mixed
is described by a wave-function(al) or a density matrix whose time
evolution is dictated by the quantum equations of motion: the
Schr\"odinger 
equation for the wave functions or the quantum Liouville equation for
a density matrix.  Consistency with the postulate 
of homogeneity and isotropy requires that the expectation value of the
energy momentum tensor must have the fluid form and in the rest frame of the fluid takes the form
$ \langle T^{\mu \nu} \rangle = \mbox{diagonal}(\rho,p,p,p) $ with
$\rho$ the energy density and $p$ the pressure. The time 
and spatial components of Einstein's equations lead to the Friedman equation
\begin{eqnarray}
&&\frac{\dot{a}^2 (t)}{a^2(t)}  =  \frac{8\pi}{3M^2_{Pl}} \rho(t)
\label{hubb} \\ 
&&2\frac{\ddot{a}(t)}{a(t)}+\frac{\dot{a}^2(t)}{a^2(t)}  =
-\frac{8\pi}{M^2_{Pl}} p(t) \label{prss} 
\end{eqnarray}
Combining these two equations one arrives at a simple and intuitive
equation which is reminiscent of the first law 
of thermodynamics:
\begin{equation}
\frac{d}{dt}(\rho a^3(t)) = -p\frac{da^3(t)}{dt} \Rightarrow \dot{\rho}+
3\frac{\dot a}{a}(\rho+p)=0 \label{firstlaw} 
\end{equation}
The alternative form shown on the right hand side of (\ref{firstlaw}) is
the {\em covariant conservation of energy}.
Since the physical volume of space is $ V_0 \;  a^3(t) $ (with $ V_0 $ the
comoving volume) the above equation is recognized as 
$ dU = -p \; dV $ which is the first law of thermodynamics for {\em
adiabatic} processes. To close the set of equations and 
obtain the dynamics we need an equation of state $ p= p(\rho) $: two very
relevant cases are: i) radiation dominated  
(RD) with $ p=\rho/3 $  and matter dominated (MD) $ p = 0 $ (dust)
Universes. In our study we will focus on the RD case.  
The equation of state for RD is that for blackbody radiation for which
the entropy is $ S= C VT^3 $ (with C a constant). 
Since $ V(t)=V_0 \;   a^3(t) $ is the physical volume, the equation
(\ref{firstlaw}) which dictates adiabatic (isoentropic) 
expansion leads to a time dependence of the temperature: $ T(t) =
T_0/a(t) $. Now the cooling is done by the expansion 
of the Universe and a phase transition will occur when the Universe
cools below the critical temperature for a given 
theory. For the GUT transition $ T_c \approx 10^{16} \, \mbox{Gev} \approx
10^{29}K $, for the EW transition 
$ T_c \approx 100 \, \mbox{Gev}\approx 10^{15}K $. Returning now back to
the large $ N $ study of the dynamics of phase transitions, 
we can include the effect of cooling by the expansion of the Universe
by replacing the time dependent mass term $ m^2(t) $ 
in (\ref{qftpotlargen}) by 
\begin{equation}
m^2(t) = m^2_0 \left[\frac{T^2(t)}{T^2_c}-1\right] ~~; ~~ T(t) =
\frac{T_i}{a(t)} \label{masstfrw} 
\end{equation}
This form is consistent with the Landau-Ginzburg description including
the time dependence of the temperature via the 
isentropic expansion of the Universe, but perhaps more importantly it
can be proven in a detailed manner from the  
self-consistent renormalization of the mass in an expanding
Universe\cite{FRW}. Thus the large $ N $ 
limit in a RD FRW cosmology will be studied by using the potential
(\ref{qftpotlargen}) but with the time dependent 
mass given by (\ref{masstfrw}).

\subsection{Large $ N $ in  Radiation (RD) and Matter (MD) dominated FRW Cosmology}
The large $ N $ limit is again implemented via the Hartree-like factorization 
(\ref{largenqft}) performing the spatial Fourier transforms of the fields
and their canonical momenta and including the proper scale factors,
the Hamiltonian now becomes\cite{FRW}
 
\begin{eqnarray}
H(t) & = &  \sum_k \left\{ \frac{1}{2a^3(t)} \;\vec{\Pi}_{\vec k}\cdot
\vec{\Pi}_{-\vec k}+ W^2_k(t) \; \vec{\Phi}_{\vec k}\cdot
\vec{\Phi}_{-\vec k} \right\} \label{hamlargenfrw}\\
W^2_k(t) & = & \frac{k^2}{a^2}+m^2(t)+\frac{\lambda}{2N} \langle
\vec{\Phi}_{\vec k}\cdot \vec{\Phi}_{-\vec k}\rangle \label{frwfreqs}
\end{eqnarray}
where now the expectation value is in terms of a {\em density matrix}
$\rho[\Phi(\vec{.}), \tilde{\Phi}(\vec{.});t]$ since we are considering
the case of a thermal ensemble as the initial state. 

We propose the following Gaussian ansatz for the functional density
matrix elements in the {Schr\"{o}dinger} representation\cite{FRW}
\begin{eqnarray}
\rho[\Phi,\tilde{\Phi},t]  =  \prod_{\vec{k}} {\cal{N}}_k(t) \exp\left\{
- \frac{A_k(t)}{2} \;\vec{\Phi}_{\vec k}\cdot \vec{\Phi}_{-\vec k}+
\frac{A^*_k(t)}{2} \;\tilde{\vec{\Phi}}_{\vec k}\cdot 
\tilde{\vec{\Phi}}_{-\vec k}+
B_k(t) \;\vec{\Phi}_{\vec k}\cdot 
\tilde{\vec{\Phi}}_{-\vec k} \right\}
      \label{densitymatrixfrw}
\end{eqnarray}
 This form of the density matrix
is dictated by the hermiticity condition $\rho^{\dagger}[\Phi,\tilde{\Phi},t] =
\rho^*[\tilde{\Phi},\Phi,t]$; as a result of this, $B_k(t)$ is real.
The kernel $B_k(t)$ determines the amount of mixing in the
density matrix, since if $B_k=0$, the density matrix corresponds to a pure
state because it is a wave functional times its complex conjugate. 
The kernels $A_k(0)~~;~~B_k(0)$ are chosen such that the initial density
matrix is thermal with a temperature $T_i > T_c$\cite{FRW}. 
Following
the same steps as in Minkowski space time, the time evolution of this 
density matrix can be found in terms of a set of mode functions $\phi_k(t)$
that obey the following equations of motion and self-consistency condition
\begin{eqnarray}
&&\ddot{\phi}_k(t)+ 3 \; \frac{\dot{a}}{a} \;\dot{\phi}_k(t)+
\left[\frac{k^2}{a^2(t)}+m^2(t)\right]\phi_k(t) = 0 \label{frweqnsofmotion} \\
&&m^2(t) = m^2_0\left[\frac{T^2_i}{T^2_ca^2(t)}-1\right]+\frac{\lambda}{2}
\int \frac{d^3k}{(2\pi)^3} \; |\phi_k(t)|^2 \;
\coth\frac{W_k(0)}{2T_i}\; .
\end{eqnarray}
This equations can be cast in a more familiar form by changing coordinates
to conformal time (see eq. (\ref{conftime})) and (conformally) rescaling the
mode functions $\phi_k(t) = f_k(\eta)/ C(\eta)$  obtaining the following
equations for the conformal time mode functions $f_k(\eta)$ in a FRW cosmology
\begin{eqnarray}
&&f''_k(\eta)+ \left[k^2+ C^2(\eta)M^2(\eta) \right]f_k(\eta) =0 
\label{conftimeqns} \\
&&M^2(eta) = m^2_0\left[\frac{T^2_i}{T^2_cC^2(\eta)}-1\right]+\frac{\lambda}{2}
\int \frac{d^3k}{(2\pi)^3} \left[ \frac{|f_k(\eta)|^2}{C^2(\eta)} \;
\coth\frac{W_k(0)}{2T_i}\right] - \frac{C''(\eta)}{C^3(\eta)} \label{conftimass}
\end{eqnarray}
where primes now refer to derivatives with respect to conformal time. For
RD  and MD FRW 
\begin{eqnarray}
&&C(\eta)= 1+ \frac{\eta}{2} ~~; ~~ C''(\eta) = 0 ~~ \mbox{for RD} \label{RDcofeta}\\
&&C(\eta)= (1+ \frac{\eta}{4})^2 ~~; ~~ C''(\eta) = 1/8 ~~ \mbox{for MD}
\label{MDcofeta}\\
\end{eqnarray}
where we have rescaled the only dimensionful variable $m_0$, length and time are now in terms of  $m^{-1}_0$. The above equations of motion  now have a
form analogous to those  in the case of Minkowski space-time. 

As the temperature falls below the critical the effective squared
mass term becomes negative and spinodal instabilities trigger the process
of phase ordering. This results in that the quantum fluctuations
quantified by $\langle \vec \Phi^2 \rangle$ grow 
exponentially. 
These spinodal instabilities make the self-consistent
field  grow at early times and tends to overcome the negative sign of
the squared mass, eventually reaching an asymptotic regime in which the
total effective mass $ M^2(\eta) $ vanishes.

 Again this behavior
determines that the fluctuations are sampling the equilibrium broken
symmetry minima of the initial potential, i.e. $\langle \vec \Phi^2
\rangle \rightarrow \frac{2Nm^2_0}{\lambda}$.

Although, just as in Minkowski space-time 
the effective mass vanishes asymptotically,  the non-equilibrium
evolution is rather {\em different}. We find numerically\cite{new}
that asymptotically the effective mass term behaves as
\begin{eqnarray}
&&C^2(\eta) M^2(\eta) \stackrel{\eta \rightarrow \infty}{\rightarrow} 
-15/4\eta^2 ~~ \mbox{for RD} \\ 
&&C^2(\eta) M^2(\eta) \stackrel{\eta \rightarrow \infty}{\rightarrow} 
-35/4\eta^2 ~~ \mbox{for MD}
\end{eqnarray}

 Fig. 6 
displays $  C^2(\eta) M^2(\eta) $ as a function of 
conformal time for the case of $ T_i/ T_c = 1.1 $ with 
$ T_c \propto m_0/\sqrt{\lambda} $\cite{FRW,boylee} for RD.

We see that at very
early time the mass is positive, reflecting 
the fact that the initial state is in equilibrium at an initial
temperature larger than the critical. As time evolves 
the temperature is  red-shifted and  cools and at some point the
phase transition occurs, when the mass vanishes 
and becomes negative.

Figure 7 displays $\frac{\lambda}{2Nm^2_0}\langle \vec{\Phi}^2
\rangle(\eta)$ vs. $\eta$  in units of 
$m^{-1}_0$ for $\frac{T_i}{T_c}=3$, $g=10^{-5}$ for an  R.D. Universe. Clearly at large times 
the non-equilibrium fluctuations probe the broken symmetry states.

This particular asymptotic behavior of the mass determines that 
 the mode functions $ f_k(\eta) $ grow as $ \eta^{5/2} $ for RD and for
$ \eta^{7/2} $  $k < 1 / \eta$ and oscillate in the form $ e^{\pm i k \eta}$ for 
$ k > 1 / \eta $. This behavior is confirmed
 numerically\cite{new}. We find both analytically and numerically\cite{new} that 
asymptotically the mode functions are of the following form  in terms of the
scaling variable $x=k\eta$
\begin{eqnarray}
&&f_k(\eta) = A \eta^{\frac{5}{2}} \;\frac{J_2(x)}{x^2}
+B \frac{x^2}{\eta^{\frac{3}{2}}} H_2(x) 
~~\mbox{for RD} \label{RDscalingFRW} \\
&&f_k(\eta) = A \eta^{\frac{7}{2}} \;\frac{J_3(x)}{x^3}
+B \frac{x^3}{\eta^{\frac{5}{2}}} H_3(x)
 ~~\mbox{for MD}
\label{MDscalingFRW} 
\end{eqnarray}
Where $ A,B $ are numerical constants and $J_{2,3}(x)$ are Bessel functions. For fixed $x$, and $\eta >>1$ the asymptotic
behavior is completely determined by the $J_{2,3}(x)$, i.e., asymptotically the solutions are of the {\em scaling form}.

Figure 8  displays $\eta^{-5}|f_k(\eta)|^2$ as a function of the
scaling variable $k\eta$ revealing the scaling behavior in RD, a similar
behavior emerges for MD\cite{new}. The absolute value of the coefficient $A$ is {\em completely fixed} by the
sum rule\cite{new} 
\begin{equation}
\frac{\lambda}{2Nm^2_0}\langle \vec{\Phi}^2
\rangle(\eta) = 1
\end{equation}
in particular this sum rule determines that $|A|^2 \propto 1/g$.

It is
remarkable that this is exactly the same scaling solution 
found in the {\em classical} non-linear sigma model in the large $ N $
limit and that describes the collapse of textures\cite{turok},
and also within the context of TDGL equations in the large $ N $ limit
applied to cosmology\cite{filipe}. However, there are important differences that are purely quantum
mechanical in origin: the Bessel functions $H_{2,3}(x)$ in the solutions (\ref{RDscalingFRW},\ref{MDscalingFRW}) 
are necessary to maintain the constancy of the Wronskian for the mode functions, the coefficients $A,B$ contain
information of the initial state. The contribution of the $H_{2,3}(x)$ breaks the scaling property of the solutions,
but they are necessary for a consistent solution for the effective mass term that falls off as $\eta^{-2}$. In particular
this behavior of the effective mass constrains the {\em non-perturbative} contribution to 
$B$ of ${\cal O}(1/\sqrt{g})$\cite{new}.

The growth of the long-wavelength modes and the oscillatory behavior
of the short wavelength modes again results in that  
the peak of the structure factor $ S(k,\eta) = |f_k(\eta)|^2 \propto C^2(\eta)\eta^3 g(k\eta)$  moves
towards longer 
wavelengths and the maximum amplitude increases.  This is the 
equivalent of coarsening and the onset of a condensate.

Although quantitatively different from Minkowsky space time, the
qualitative features are similar. Asymptotically the non-equilibrium
dynamics results in the formation of a non-perturbative 
condensate of long-wavelength Goldstone bosons. We can now compute the
pair correlation function $C(r,\eta)$ from the mode functions solutions to
(\ref{conftimeqns}) and find that it is cutoff by causality at
$r=2\eta$. The correlation function computed with the mode functions in the asymptotic regime agrees perfectly with that computed from the asymptotic
form given by (\ref{RDscalingFRW}).  The correlation function is depicted in Fig. 9  for two different (conformal) times.

 The scaling form of the pair
correlation function is
$$
C(r,\eta) \propto \eta^2 \; \chi(r/2\eta)
$$
where $ \chi(x) $ is a hump-shaped function as shown in fig. 9. 

Clearly a {\em dynamical} length scale $L(\eta) = \eta$ emerges as a consequence of causality, much in the
same manner as in Minkowsky space time. The {\em physical} dynamical correlation length is therefore given by
$\xi_{phys}(\eta) = C(\eta) L(\eta) = d_H(t)$,
that is the correlated domains grow again at the speed of light and their size is given by the causal horizon. 
The interpretation of this phenomenon is that within one causal horizon there is one correlated domain, inside which
the mean square root fluctuation
of the field is approximately the value of the equilibrium minima of the tree level potential, this is clearly
consistent with Kibble's original observation\cite{kibble,kibble2}. Inside this domain there
is a non-perturbative condensate of Goldstone bosons\cite{new}. 

There are remarkable consequences of the scaling solution\cite{new}:

\begin{itemize}
\item{When the scale factor $C(\eta) \propto \eta^{\alpha}$ ($\alpha =1,2$ for RD and MD respectively) the asymptotic
solution is {\em universal} in the sense that it {\em does not } depend on the past history of the background metric,
the coefficient $A$ is completely determined by the sum rule an is insensitive to the past evolution of the scale
factor. This is important in the case of a transition between RD and MD, if the two regimes are well separated in
conformal time, the asymptotic solution in the RD and MD eras is independent of the transition\cite{new}.}

\item{We find\cite{new} that the fluid that results from the fluctuations obeys the {\em same equation of state of the
background fluid}. In particular we find $p= \frac{e}{3} \left(\frac{2}{\alpha}-1\right)$, i.e. 
$p=e/3$ for $\alpha=1$(RD)  and $p=0$ for $\alpha=2$ (MD).}

\end{itemize}

Thus we have seen that the phenomenon of scaling, coarsening and the onset of condensates during the non-equilibrium dynamics of phase ordering is a
{\em universal} feature of the process of phase ordering. The non-perturbative large N limit has allowed a clear comparison between the phenomenological description in condensed matter based on the TDGL, and the microscopic quantum field theoretical description in Minkowski and FRW space-times.

\section{Conclusions and looking ahead} 

In this lectures we have discussed the multidisciplinary nature of
the problem of phase ordering kinetics and 
non-equilibrium aspects of symmetry breaking. Main ideas from
condensed matter were discussed and presented in 
a simple but hopefully illuminating framework and applied to the rather different realm of phase transitions in quantum field theory as needed to understand cosmology and particle physics. 
In particular we have emphasized {\em robust} features of the process of phase ordering kinetics: early stages dominated by spinodal instabilities and the growth of correlated regions, the emergence of a dynamical correlation length that determines the size of the correlated regions as a function of time and  {\em dynamical scaling} at long times. The phenomenon of coarsening is a result of this scaling behavior and is reflected in that the peak in the power spectrum moves towards longer wavelengths, and asymptotically long times results in a ``Bragg peak'' that signals the onset of macroscopic ordered phases and condensates. The study of condensed matter systems was in terms of the phenomenologically succesful Time Dependent Landau Ginzburg theory which is purely dissipative and for which there is no first principles derivation from a microscopic theory in general. 

We then passed onto the study of the dynamical evolution out of equilibrium in quantum field theories both in Minkowsky and FRW space-times by providing a consistent {\em non-perturbative} framework to study the time evolution of an initially prepared density matrix.  
 
The large $ N $ approximation has provided a bridge that
allows to cross from one field to another and borrow many of the ideas that had been tested both theoretically and
experimentally in condensed matter physics. There are, however, major differences between the condensed matter and
particle physics-cosmology applications that require a very careful treatment of the quantum field theory that cannot
be replaced by simple arguments. The large $ N $ approximation in field theory provides a robust, consistent non-perturbative
framework that allows the study of phase ordering kinetics and dynamics of symmetry breaking in a controlled and consistently implementable framework, it is renormalizable, respects all symmetries and can be improved in a well defined
manner. This scheme extracts cleanly the non-perturbative behavior, the quantum to classical transition and allows to
quantify in a well defined manner the emergence of classical stochastic behavior arising from non-perturbative physics. 
The emergence of scaling and a dynamical correlation length, coarsening and the onset of non-perturbative condensates are robust features of the dynamics and the Kibble-Zurek
scenario describes fairly well the general features of the dynamics, albeit the details require careful study, both
analytically and numerically. 

We have emphasized that this study has very definite potential experimental implications, in QCD if the chiral phase transition occurs out of equilibrium in ultrarelativistic heavy ion collisions leads to the possibility of formation of disoriented chiral condensates that are described in the same manner as ordering domains in condensed matter. These
condensates have a very distinct hallmark in that they lead to a very different ratio of neutral to charged pions, this property can be measured on an event by event basis with the detectors at the  forthcoming heavy ion colliders. 

In cosmology the process of formation of ordered regions that grow after
a rapid phase transition, the emergence of scaling and a dynamical length
scale and coarsening of these domains lead to a definite prediction of a
``red'' power spectrum on scales that have re-enterd the causal horizon
right after recombination. These are the scales that contribute to the temperature anisotropies measured by COBE and the forthcoming cosmological experiments.  
Therefore the study of the dynamics of symmetry breaking out of equilibrium in quantum field theory directly bears on experimental possibilities in a wide range of energies both in accelerator and cosmological experiments and is therefore an endeavour that must be pursued vigorously. 

Of course this is just the beginning, we expect a wealth of important phenomena to be revealed beyond the large $N$, such
as the approach to equilibrium, the emergence of other time scales associated with a hydrodynamic description of the
evolution at late times and a more careful understanding of the reheating process and its influence on cosmological
observables. Although within very few years the wealth of
observational data will provide a more clear picture of the
cosmological fluctuations, it is clear that the program that pursues
a fundamental understanding of the underlying physical mechanisms will continue
seeking to provide  a consistent microscopic description of the
dynamics of particle physics and  cosmological phase transitions.  

\section{Acknowledgements:} 
 D. B. thanks T. Kibble, W. Zurek and R. Durrer for illuminating 
conversations, the N.S.F for
partial support through grant awards: PHY-9605186 and INT-9815064 and LPTHE (University of Paris VI and VII) for warm
hospitality, H. J. de Vega thanks the Dept. of Physics at the Univ. of Pittsburgh for hospitality.  We thank NATO for partial support.





\begin{figure}
\centerline{ \epsfig{file=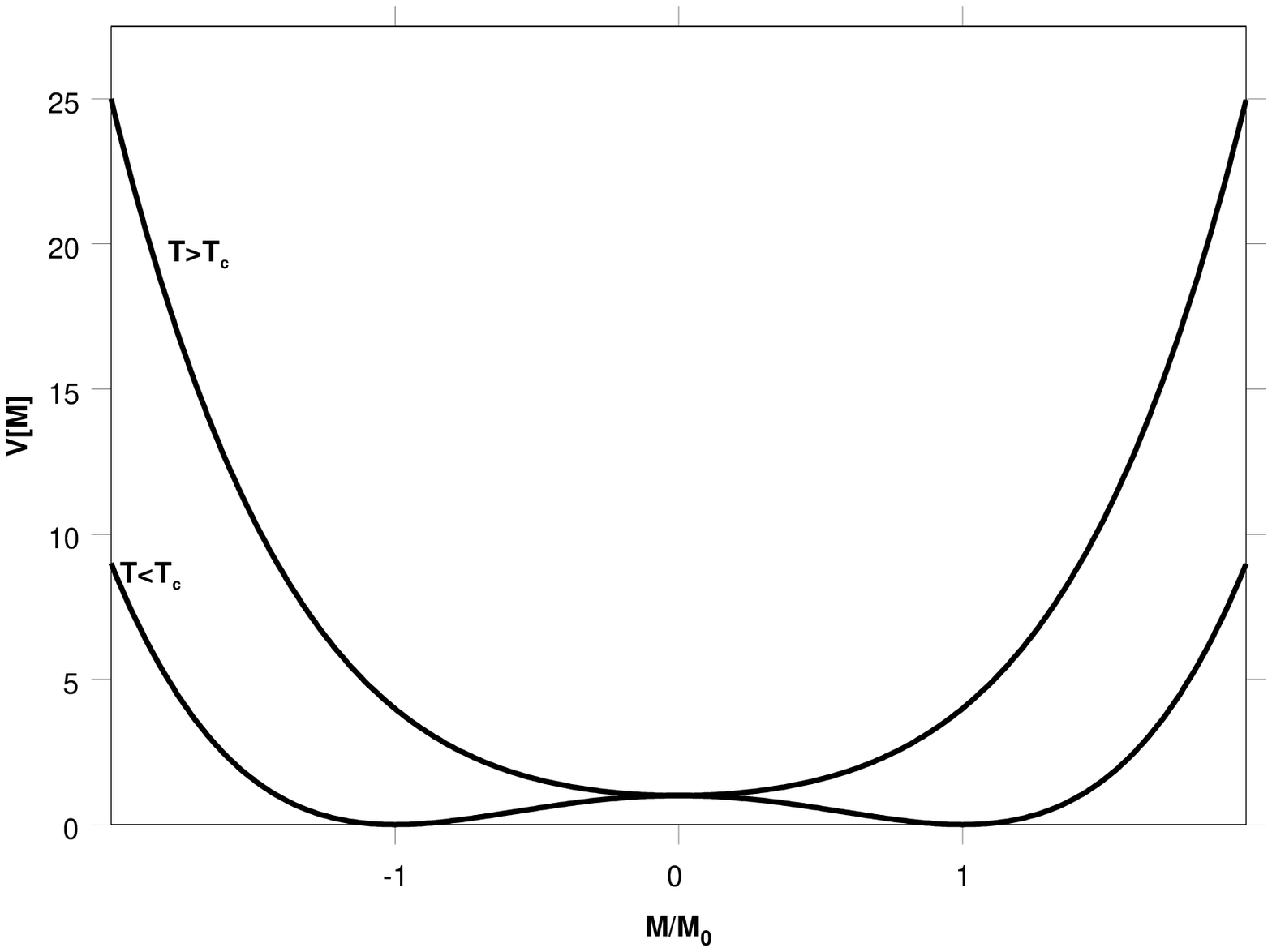,width=7in,height=8in}}
\caption{$V[M]$ vs. $M$, for $T>T_c$ and $T<T_c$ \label{fig1}}
\end{figure}


\begin{figure}
\centerline{ \epsfig{file=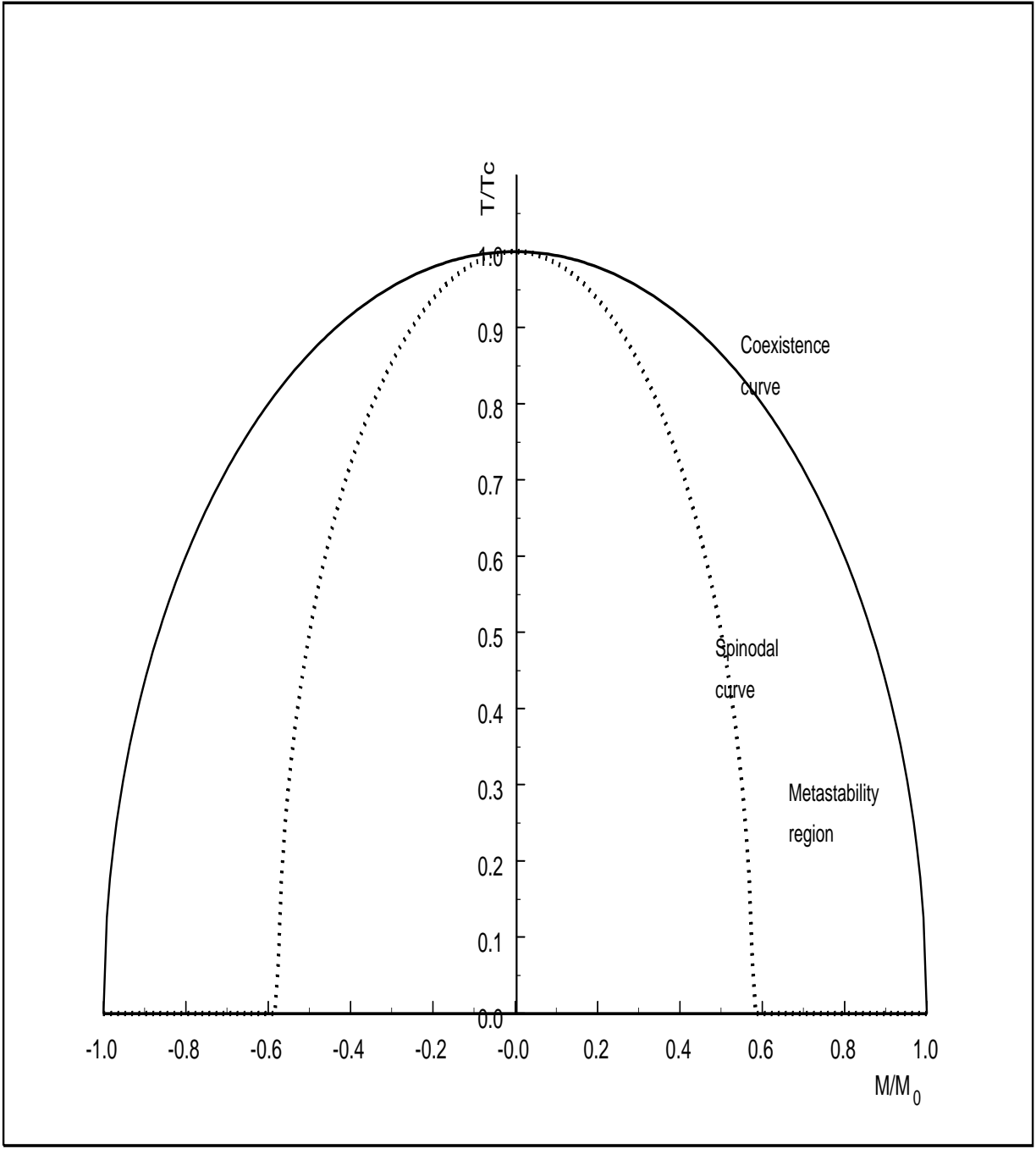,width=7in,height=8in}}
\caption{Classical spinodal and coexistence curves for the potential $V[M]$ in (\ref{freenergy}) \label{fig2}}
\end{figure}


\begin{figure}
\centerline{ \epsfig{file=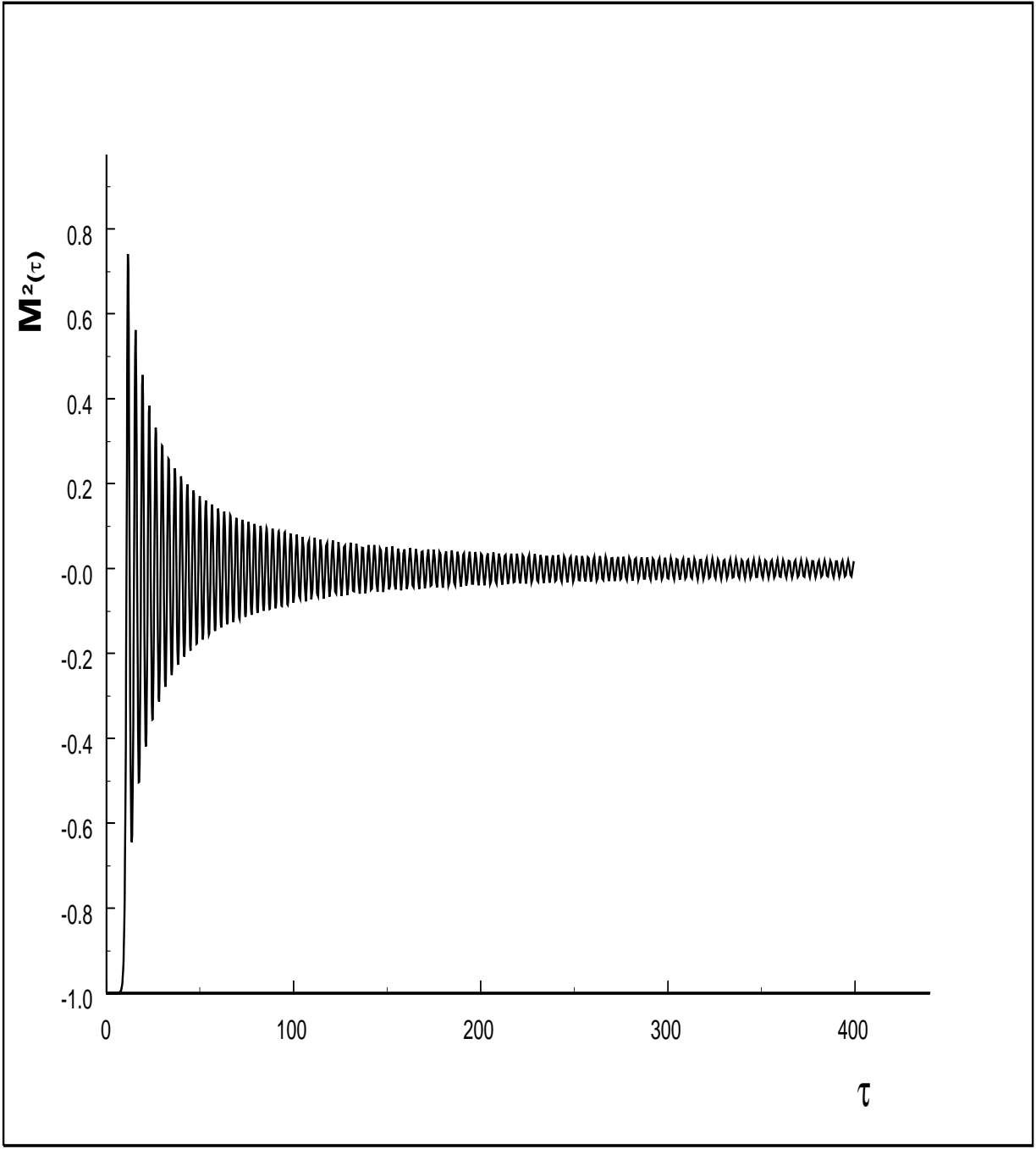,width=7in,height=8in}}
\caption{${\cal M}^2(\tau)$ vs. $\tau$, $g=10^{-7}$ \label{fig3}}
\end{figure}



\begin{figure}
\centerline{ \epsfig{file=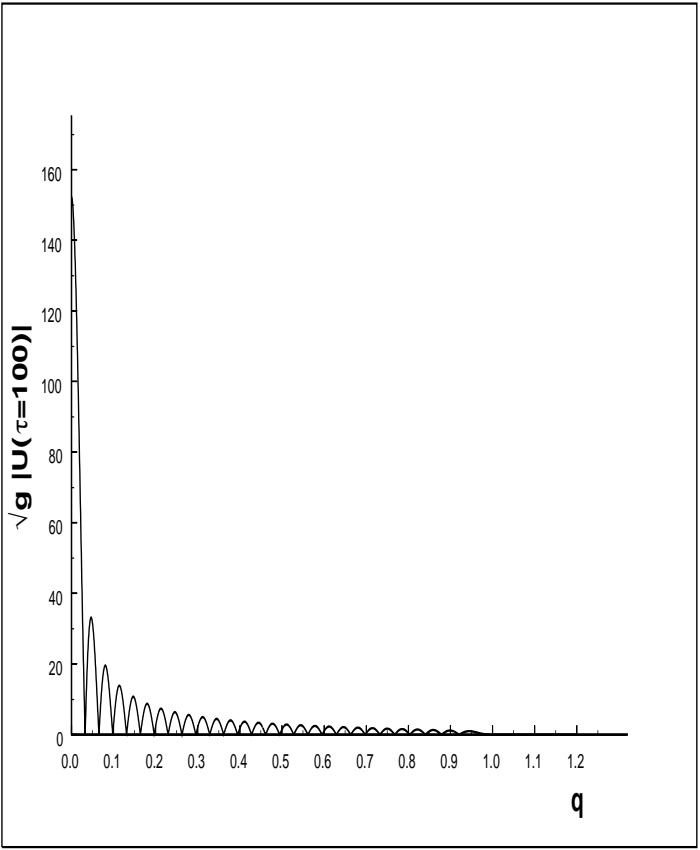,width=4in,height=5in}  \epsfig{file=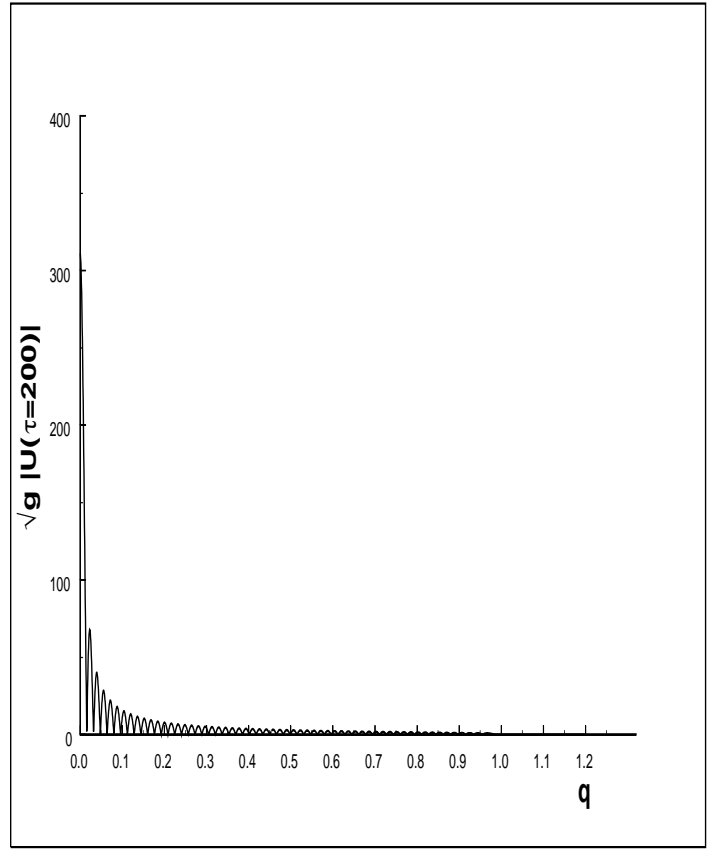,width=4in,height=5in} }
\caption{$g|\phi_k(\tau=100,200)|^2$ vs. $q=k/|m_{0}|$, $g=10^{-7}$ \label{fig4}}
\end{figure}



\begin{figure}
\epsfig{file=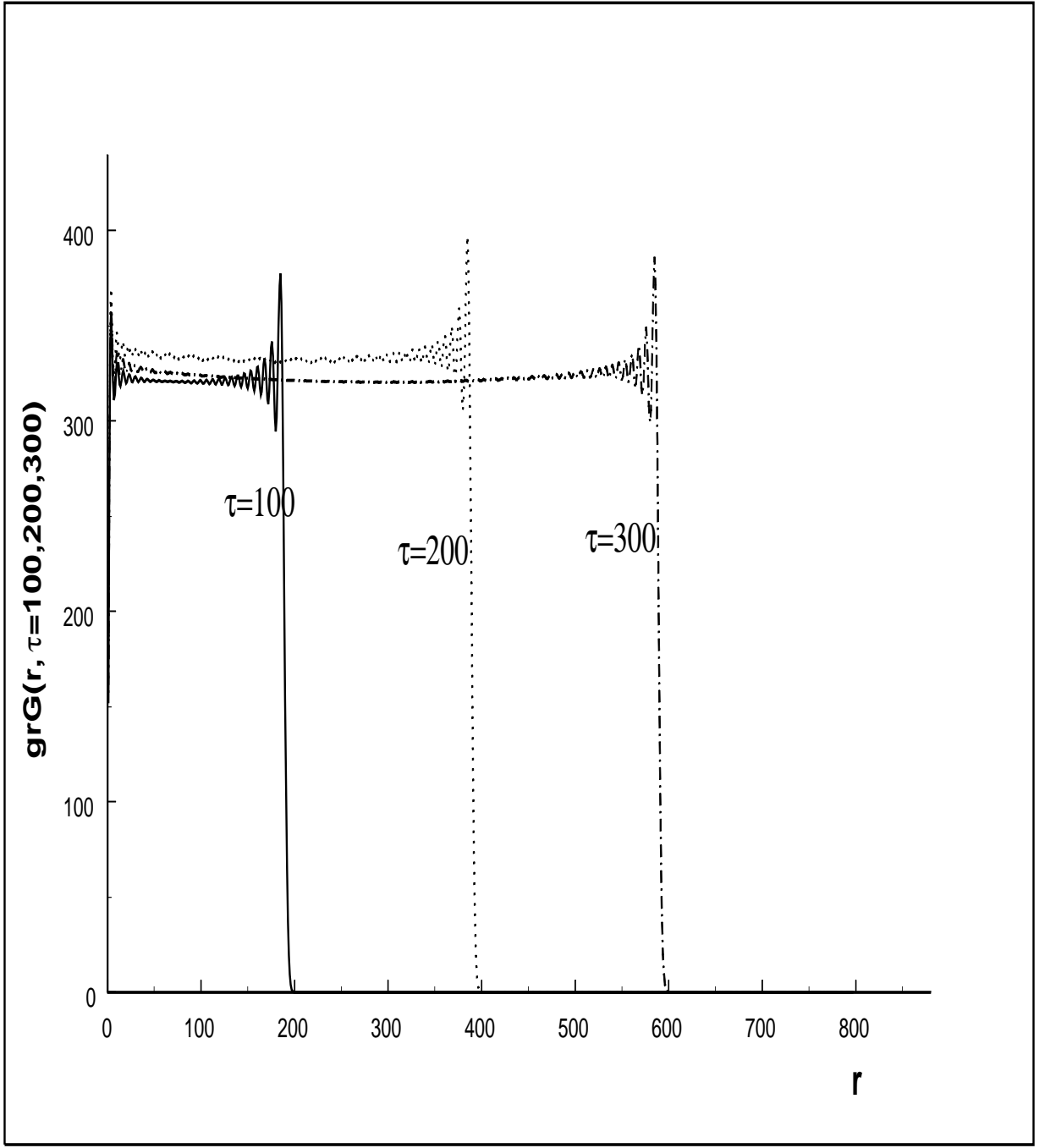,width=7in,height=8in}
\caption{$ g r\, C(r,\tau) $ vs $ r/|m_{0}| $ for $ t/|m_0| = 100, \;
 200, \; 300 $  for $ g = 10^{-7} $.
 \label{fig5}}
\end{figure}



\begin{figure}
{\epsfig{file=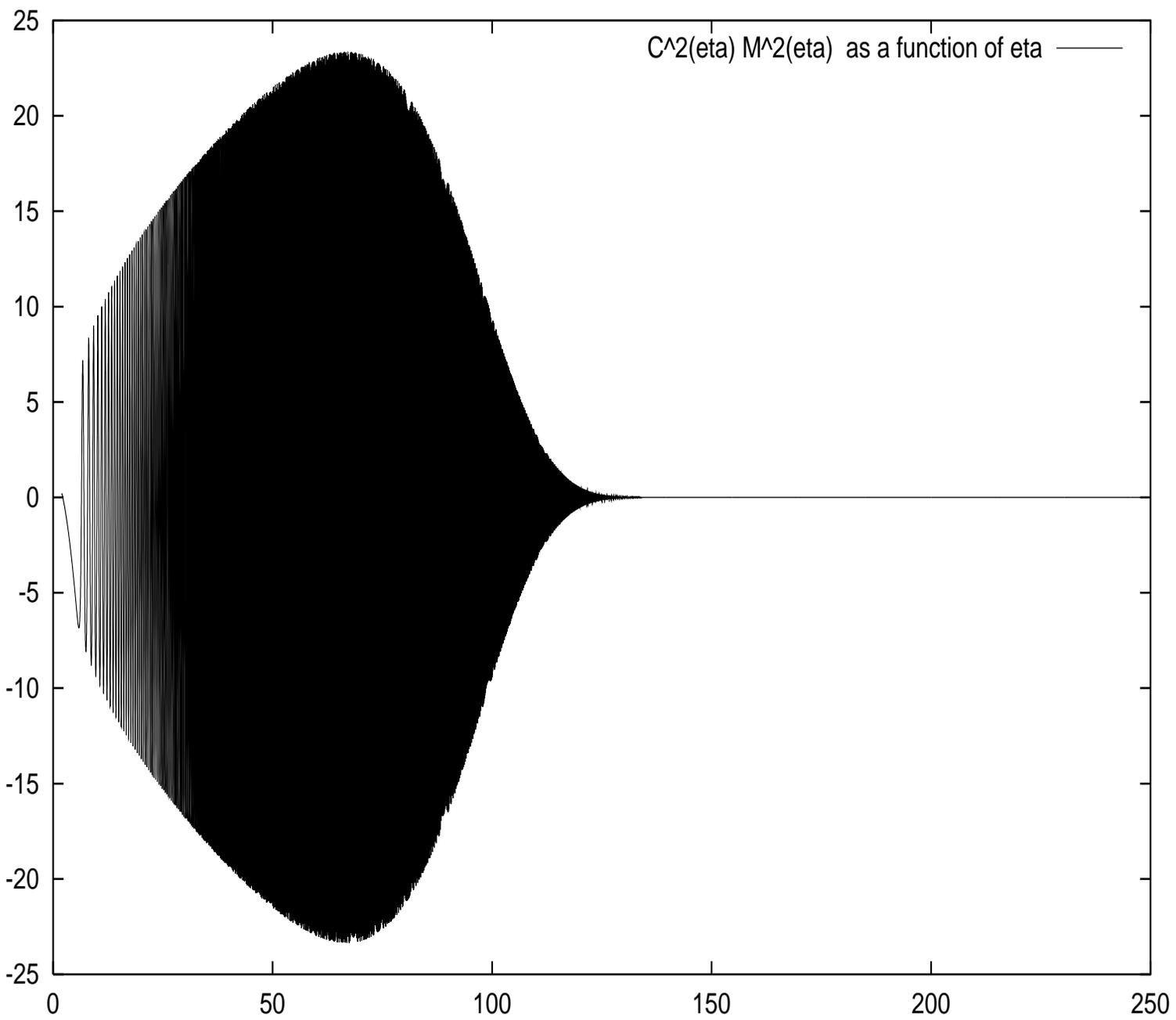,width=7in,height=8in}}
\caption{ $C^2(\eta) M^2(\eta)$ vs. $\eta$(conformal time in units of
$m^{-1}_0$) for $\frac{T_i}{T_c}=3$, $g=10^{-5}$. R.D. Universe.
\label{Fig6} } 
\end{figure}



\begin{figure}
{\epsfig{file=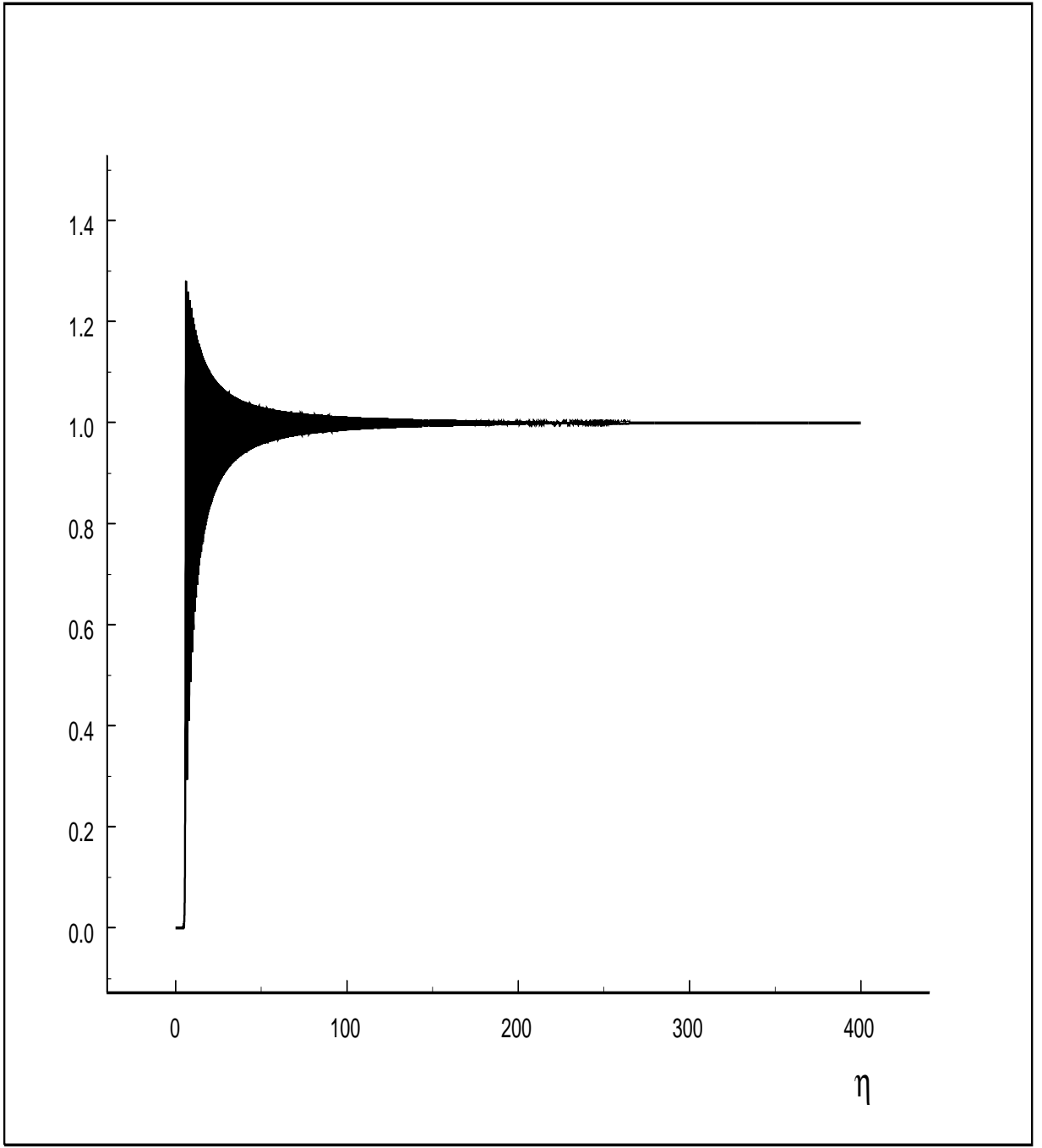,width=7in,height=8in}}
\caption{ $\frac{\lambda}{2Nm^2_0}\langle \vec{\Phi}^2
\rangle(\eta)$ vs. $\eta$ (conformal time in units of 
$m^{-1}_0$) for $\frac{T_i}{T_c}=3$, $g=10^{-5}$.  R.D. Universe.\label{Fig7} } 
\end{figure}



\begin{figure}
{\epsfig{file=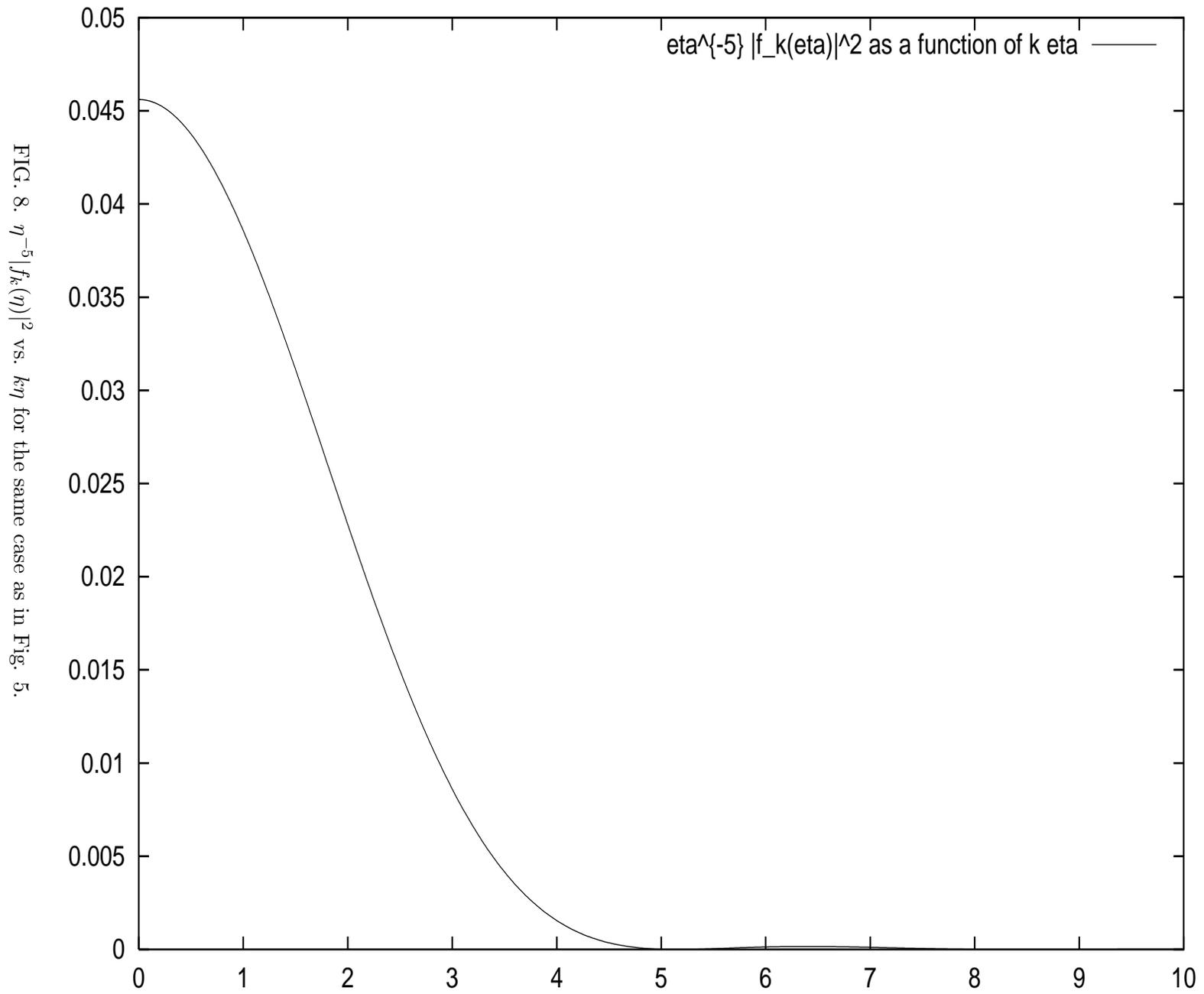,width=7in,height=8in}}
\caption{$\eta^{-5}|f_k(\eta)|^2$ vs. $k\eta$ for the same case as in Fig. 5.\label{Fig8}}
\end{figure}



\begin{figure}
{\epsfig{file=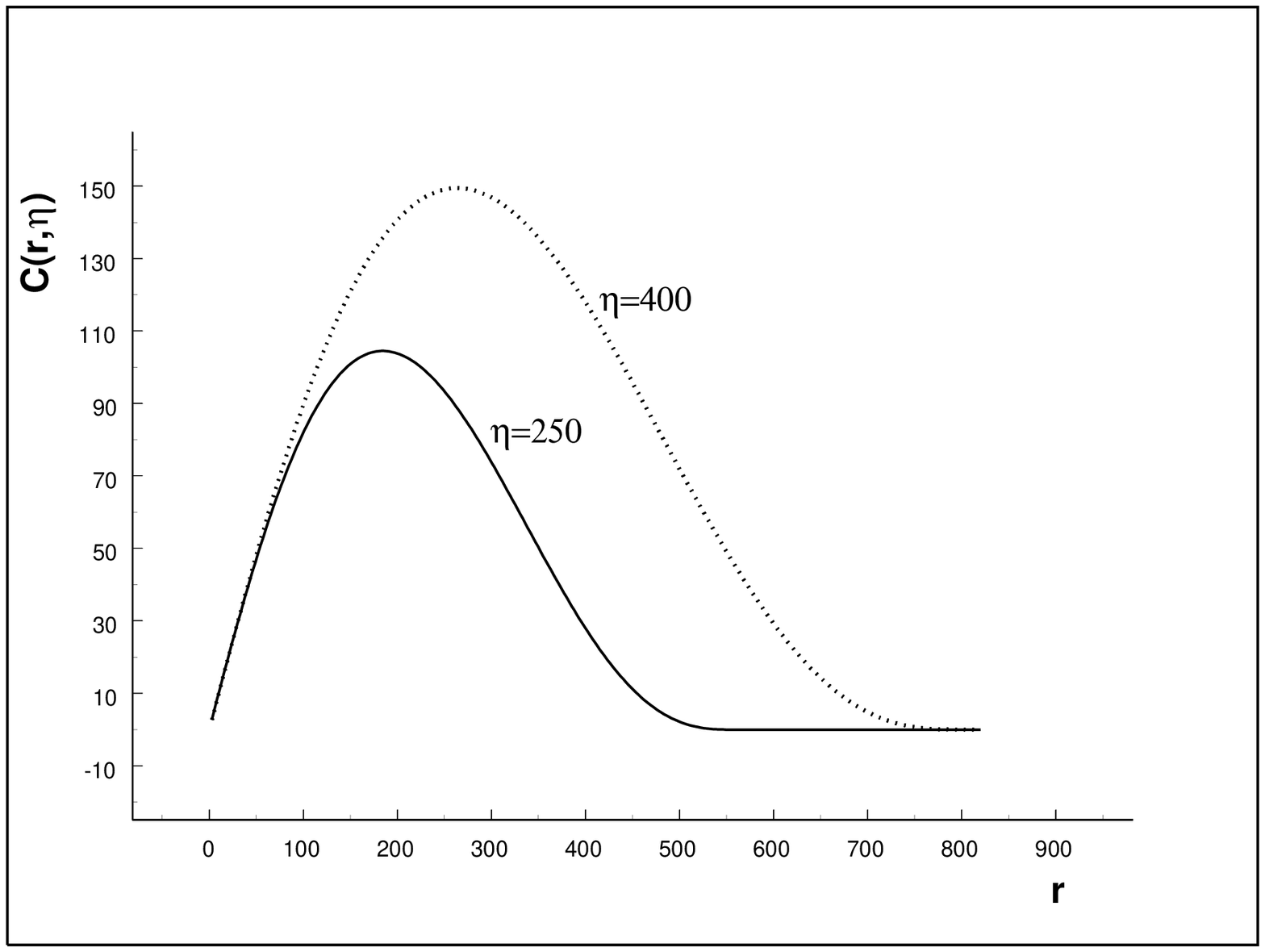,width=7in,height=8in}}
\caption{$C(r,\eta)$ vs. $r$ for $\eta=250,400$ (in units of 
$m^{-1}_0$) for $\frac{T_i}{T_c}=3$, $g=10^{-5}$.  R.D. FRW Universe. \label{Fig9}}
\end{figure}



\begin{thebibliography}{99}

\bibitem{kibble} T. W. B. Kibble, J. Phys. A 9, 1387 (1976), and contribution to these proceedings.

\bibitem{kibble2} M. B. Hindmarsh and T.W.B. Kibble,
  Rep. Prog. Phys. {\bf 58}:477 (1995). 


\bibitem{vilen} A. Vilenkin and E.P.S. Shellard, `Cosmic Strings and
  other Topological Defects',  Cambridge Monographs on Math. Phys. (Cambridge
  Univ. Press, 1994). 

\bibitem{durrer} For a comprehensive review of the status of theory
and experiment see: Proceedings of the `D. Chalonge' School in
Astrofundamental Physics at Erice, edited by N. S\'anchez and
A. Zichichi, 1996 World Scientific publisher and 1997, Kluwer Academic
publishers. In particular the contributions by G. Smoot, A. N. Lasenby
and A. Szalay.,  

R. Durrer, M. Kunz and A. Melchiorri
astro-ph/9811174, M. Kunz and R. Durrer, Phys. Rev. D55, R4516 (1997).  


\bibitem{rajagopal} See for example: K. Rajagopal in `Quark Gluon
Plasma 2', (Ed. R. C. Hwa, World Scientific, 1995). 


\bibitem{goldenfeld} For a thorough discussion of phase transitions see:
N. Goldenfeld, `Lectures on Phase Transitions and the Renormalization Group', (Addison-Wesley, 1992). 

 

\bibitem{bray} A. J. Bray, Adv. Phys. {\bf 43}, 357 (1994). 

\bibitem{langer} J. S. Langer in `Solids far from Equilibrium',
Ed. C. Godr\`eche, (Cambridge Univ. Press 1992); 
J. S. Langer in `Far from Equilibrium Phase Transitions',
Ed. L. Garrido, (Springer-Verlag, 1988); J. S. Langer 
in `Fluctuations, Instabilities and Phase Transitions', Ed. T. Riste,
Nato Advanced Study Institute, Geilo Norway, 1975 (Plenum, 1975).

\bibitem{mazenko} G. Mazenko in in `Far from Equilibrium Phase
Transitions', Ed. L. Garrido, (Springer-Verlag, 1988).   

\bibitem{marco} C. Castellano and M. Zannetti, cond-mat/9807242;
C. Castellano, F. Corberi and M. Zannetti, Phys. Rev. E56, 4973
(1997); F. Corberi, A. Coniglio and M. Zannetti, Phys. Rev. E51, 5469 (1995). 

\bibitem{zurek} W. H. Zurek, Nature 317, 505 (1985); Acta Physica 
Polonica B24, 1301 (1993); Phys. Rep. 276, 4 (1996).  

\bibitem{goldburg} W. I. Goldburg and J. S. Huang, in `Fluctuations, Instabilities and Phase Transitions', Ed. T. Riste, Nato Advanced Study Institute, Geilo
Norway, 1975 (Plenum, 1975); J. S. Huang, W. I. Goldburg and
M. R. Moldover, Phys. Rev. Lett. {\bf 34}, 639 (1975).  

\bibitem{gill} For a  nicely written recent review on the dynamics
of phase transition see: A. Gill, `Contemporary Physics', vol. 39, number 1,
pages 13-47 (1998). 

\bibitem{turokexp} I. Chuang, R. Durrer, N. Turok and B. Yurke, Phys. Rev.
Lett. 66, 2472 (1990).

\bibitem{bowickexp} M. Bowick, L. Chandar, E. Schiff and A. Srivastava,
Science 263, 943 (1994). 

\bibitem{lancaster} P. C. Hendry, N. S. Lawson, R. A. M. Lee, P. V. E. McClintock and C. D. H. Williams, Nature, 368, 315 (1994). 

\bibitem{ruutu} V.M.H. Ruutu et. al., `Big Bang simulation in superfluid $^3\mbox{He}$-B, vortex nucleation in neutron irradiated superflow'',
cond-mat/9512117. 

\bibitem{bunkov} Y. M. Bunkov and O. D. Timofeevskaya, `Cosmological scenario for A-B phase transition in superfluid $^3\mbox{He}$', cond-mat/9706004.  

\bibitem{QCD} For recent reviews on the QCD phase transitions and aspects
of relativistic heavy ion collisions see for example:  J. W. Harris and B. Muller, Annu. Rev. Nucl. Part. Sci.
{ 46}, 71 (1996). B. Muller in {\em Particle Production in Highly
Excited Matter}, 
Eds. H.H. Gutbrod and J. Rafelski, NATO ASI series B, vol. 303
(1993). B. Muller, {\em The Physics 
of the Quark Gluon Plasma} Lecture Notes in Physics, Vol. 225 (Springer-Verlag,
Berlin, 
Heidelberg, 1985);  K. Rajagopal in ``Quark-Gluon Plasma 2'', Ed. by R. C. Hwa (World Scientific, Singapore) (1995); H. Meyer-Ortmanns, Rev. of Mod. Phys. {\bf 68}, 473 (1996). C-Y Wong, `Introduction to High-Energy Heavy Ion Collisions', (World Scientific, 1994). 

\bibitem{bjorkenhydro} J. D. Bjorken, Phys. Rev. D27, 140 (1982).

\bibitem{raja} K. Rajagopal and F. Wilczek, Nucl. Phys. B399, 395 (1993);
K. Rajagopal and F. Wilczek, Nucl. Phys. B404, 577 (1993). 



\bibitem{dcc} A. A. Anselm and M. G. Ryskin, Phys. Lett. {\bf B266}, (1991)
482; J. D. Bjorken, K. L. Kowalski and C. C. Taylor, SLAC Report No. SLAC-PUB-6109 (unpublished); J. - P. Blaizot and A. Krzywicki, Phys. Rev. {\bf D46}, 1992 (246);  J. D. Bjorken, Int. J. Mod. Phys. {\bf A7}, (1992) 4189; J. D. Bjorken, Acta Physica Polonica {\bf B23}, (1992) 561;
K. L. Kowalski and C. C. Taylor, ``Disoriented Chiral
Condensate: A White Paper for the Full Acceptance Detector'' CWRU report 92-
he-ph/9211282 (unpublished); J. D. Bjorken, K.L. Kowalski and C. C. Taylor, ``Baked
Alaska'', Proceedings of Les Rencontres de Physique del Valle d'Aoste, La
Thuile (1993);  (SLAC PUB 6109).  G. Amelino-Camelia, J. D. Bjorken, S. E. Larsson, Phys.Rev. D56 (1997) 6942;  J. D. Bjorken, Acta Phys.Polon. B28 (1997) 2773; A. Anselm, Phys. Lett. B217, 169 (1989). 

\bibitem{centauro} L. T. Baradzei et. al. Nucl. Phys. {\bf B370}, (1992)
365. 


\bibitem{moredcc} S. Gavin, A. Gocksch and R. D. Pisarski, Phys. Rev. Lett, {\bf 72}, 2143 (1994); S. Gavin and B. Muller, Phys. Lett. B329, 486 (1994);
 Z. Huang and X.-N. Wang, Phys. Rev. {\bf D49}, 4335 (1994); Z. Huang,
M. Suzuki and X-N. Wang, Phys. Rev. {\bf D50}, 2277 (1994); 
 Z. Huang and M. Suzuki, Phys. Rev. D53, 891 (1996); M. Asakawa, Z. Huang and X. N. Wang, Phys. Rev. Lett. 74, 3126 (1995);  J. Randrup, Nucl.Phys. {\bf A616} (1997) 531; J. Randrup, Phys.Rev.Lett. {\bf 77} (1996) 1226.

\bibitem{boydcc} D. Boyanovsky, H. J. de Vega and R. Holman, Phys. Rev. {\bf D51}, (1995) 734; F. Cooper,  Y. Kluger, E. Mottola and J. P.
Paz. Phys. Rev. {\bf D51}, (1995) 2377 . Y. Kluger, F. Cooper, E. Mottola, J. P. Paz and A.
Kovner, Nucl. Phys. {\bf A590},(1995) 581. 

\bibitem{wa98} WA98 Collaboration, (M. M. Aggarwal et. al.) Phys. Lett. B
{\bf B420}, (1998)  169. 
%

\bibitem{fermilab} J. Streets, ``Preliminary results from a search for 
Disoriented Chiral Condensate at Minimax'', hep-ex/9608012; T. C. Brooks et. al. Phys. Rev. D55, (1997), 5667; M. E. Convery, ``A disoriented chiral 
condensate search at the Fermilab Tevatron'', hep-ex/9801020.


\bibitem{rhic} see the RHIC project page with detailed description of the physics capabilities of STAR and PHENIX at
$\mbox{http://www.rhic.bnl.gov}$

\bibitem{castor} see the Castor project page at the Alice web page, $\mbox{http://www1.cern.ch/ALICE/projects.html}$.


\bibitem{kolb}For   thorough reviews of standard and inflationary 
cosmology see: E. W. Kolb and M. S. Turner, {\em The Early Universe}
(Addison Wesley, Redwood City, C.A. 1990). A. Linde, {\em Particle
Physics and Inflationary Cosmology}, (Harwood Academic Pub. Switzerland,
1990). R. Brandenberger, Rev. of Mod. Phys. 57,1 (1985); Int. J. Mod. Phys.
A2, 77 (1987). 

\bibitem{turner} For more recent reviews see: M. S. Turner,
astro-ph-9703197;astro-ph-9703196;astro-ph-9703174;astro-ph-9703161;
astro-ph-9704062.

\bibitem{inflation} D. Boyanovsky, D. Cormier, H. J. de Vega, R. Holman and S. P. Kumar,Phys. Rev. D. 57,2166 (1998); D. Boyanovsky, D. Cormier, H. J. de Vega and R. Holman,
Phys.Rev. D55 (1997) 3373. 







\bibitem{turok} N. Turok and D. N. Spergel, Phys. Rev. Lett. 66, 3093
(1991); D. N. Spergel, N. Turok, W. H. Press and B. S. Ryden,
Phys. Rev. D43, 1038 (1991).  

\bibitem{filipe} J. A. N. Filipe and A. J. Bray, Phys. Rev. E50, 2523
(1994); J. A. N. Filipe, (Ph. D. Thesis, 1994, unpublished). 

\bibitem{FRW} D. Boyanovsky, H. J. de Vega and R. Holman, 
Phys. Rev.  {\bf D 49}, 2769 (1994); 
D. Boyanovsky, D. Cormier, H. J. de Vega, R. Holman et S. Prem
Kumar, Phys. Rev.  {\bf D57}, 2166, (1998),
(and references therein).

\bibitem{nuestros} 
D. Boyanovsky, H. J. de Vega, R. Holman, D.-S. Lee and A. Singh, 
Phys. Rev. {\bf D51}, 4419 (1995). 
D. Boyanovsky, H. J. de Vega and R. Holman, Proceedings of
the Second Paris Cosmology Colloquium, Observatoire de Paris, June 1994,
pp. 127-215, H. J. de Vega and N. S\'anchez, Editors (World
Scientific, 1995); Advances in Astrofundamental Physics, Erice
Chalonge School, N. S\'anchez and A. Zichichi Editors, (World
Scientific, 1995).  D. Boyanovsky, H. J. de Vega, R. Holman and J. Salgado,
Phys. Rev. {\bf D54}, 7570 (1996);  D. Boyanovsky, D. Cormier,
 H. J. de Vega, R. Holman, A. Singh, M. Srednicki; Phys. Rev. D56
 (1997) 1939.  
D. Boyanovsky, H. J. de Vega and R. Holman, 
 Vth. Erice Chalonge School, Current Topics in Astrofundamental
 Physics, N. S\'anchez and A. Zichichi Editors, World Scientific,
 1996, p. 183-270. 
D. Boyanovsky, M. D'Attanasio,
H. J. de Vega, R. Holman and D. S. Lee, 
Phys. Rev. {\bf D52}, 6805 (1995).  D. Boyanovsky, H. J. de Vega,
R. Holman and J. Salgado, Phys. Rev. {\bf D57}, 7388 (1998).

\bibitem{noscorre} D. Boyanovsky, H. J. de Vega, R. Holman and
J. Salgado, hep-ph/9811273, to appear in Phys. Rev. {\bf D}. 

\bibitem{losalamos}  F. Cooper, S. Habib, Y. Kluger, E. Mottola,
 Phys.Rev. D55 (1997), 6471. F. Cooper, S. Habib, Y. Kluger,
E. Mottola, J. P. Paz, P. R. Anderson, 
Phys. Rev. {\bf D50}, 2848 (1994). 
F. Cooper, Y. Kluger, E. Mottola, J. P. Paz, Phys. Rev. {\bf D51},
2377 (1995); F. Cooper and E. Mottola, Mod. Phys. Lett. A 2, 635 (1987);
F. Cooper and E. Mottola, Phys. Rev. D36, 3114
(1987); F. Cooper, S.-Y. Pi and P. N. Stancioff,
Phys. Rev. D34, 3831 (1986). 

\bibitem{new} D. Boyanovsky and H. J. de Vega, ``Dynamics of Symmetry Breaking in FRW Cosmologies'' (in progress). 

\bibitem{cugliandolo} L. F. Cugliandolo and D. S. Dean, J. Phys. A28, 4213 (1995); {\em ibid} L453, (1995); 
L. F. Cugliandolo, J. Kurchan and G. Parisi, J. Physique (France) 4,
1641 (1994). 

\bibitem{broken} See D. Boyanovsky, R. Holman and H. J. de Vega in \cite{boydcc}, and the first reference in \cite{losalamos}.


\bibitem{bowick} Relaxing the assumption of an instantaneous quench
and allowing for a time dependence of the cooling mechanism has been
recently studied by  M. Bowick and A Momen, hep-ph/9803284.  


\bibitem{erickwu}  E. J. Weinberg and A. Wu, Phys. Rev. D36, 2474 (1987); A. Guth and S.-Y. Pi, Phys. Rev. D32, 1899 (1985).

\bibitem{boyvega}D. Boyanovsky and H. J. de Vega, Phys. Rev. {\bf D47}, 2343 (1993); D. Boyanovsky Phys. Rev.  E48, 767 (1993). 

\bibitem{boylee} D. Boyanovsky, D.-S. Lee and A. Singh, Phys. Rev. D48, 800 (1993). 


\bibitem{rivers}  G. Karra and  R.J.Rivers, Phys.Lett. B414 (1997), 28;
 R.J.Rivers, 3rd. Colloque Cosmologie, Observatoire de Paris, June
 1995, p. 341 in the Proceedings edited by H J de Vega and
 N. S\'anchez, World Scientific. 
A.J. Gill and R.J. Rivers, Phys.Rev. D51 (1995), 6949; 
G.J. Cheetham, E.J. Copeland, T.S. Evans, R.J. Rivers, 
 Phys.Rev.D47 (1993),5316. 

\bibitem{beilok} `Defect Formation and Critical Dynamics in the Early
Universe', 
G. J. Stephens, E. A. Calzetta, B. L. Hu, S. A. Ramsey, gr-qc/9808059 (1998).
  `Counting Defects in an Instantaneous Quench', D. Ibaceta and E.
Calzetta, hep-ph/9810301 (1998).

 



 

\end{thebibliography}
\end{document}